\documentclass[12pt]{iopart}
\usepackage{cite}
\usepackage[T1]{fontenc}
\usepackage[utf8]{inputenc}
\usepackage{lmodern}
\usepackage{verbatim}
\usepackage{microtype}
\usepackage{times}
\usepackage[varg]{txfonts}

\usepackage[dvipsnames, usenames]{xcolor}
\definecolor{linkcolor}{rgb}{0.0,0.3,0.5}
\usepackage[hypertexnames=false, unicode, colorlinks=true, linkcolor=linkcolor,
citecolor=linkcolor, filecolor=linkcolor,urlcolor=linkcolor,
pdfusetitle]{hyperref}

\usepackage[all]{hypcap}
\usepackage{graphicx}
\usepackage{xspace}
\usepackage{amssymb}
\usepackage{iopams}
\usepackage[normalem]{ulem}
\usepackage{bm}

\usepackage{enumitem}

\usepackage[english]{babel}
\usepackage{orcidlink}

\usepackage{tikz}
\usetikzlibrary{shapes.geometric, arrows, calc}

\graphicspath{%
  {figs/}%
}

\DeclareMathAlphabet{\mathpzc}{OT1}{pzc}{m}{it}

\newcommand{\roughly}{\mathchar"5218\relax\,}

\newcommand{\h}{\mathpzc{h}}
\newcommand{\hlm}{\mathpzc{h}_{\ell, m}}

\newcommand{\ylm}{{}_{-2}Y_{\ell, m}}
\newcommand{\philm}{\phi_{\ell, m}}
\newcommand{\amplm}{A_{\ell, m}}

\newcommand{\dd}{\mathrm{d}}

\newcommand{\egw}{e_{\mathrm{gw}}}
\newcommand{\egwAS}{e^{\mathrm{as}}_{\mathrm{gw}}}

\newcommand{\lgw}{l_{\mathrm{gw}}}
\newcommand{\omegaA}{\omega^{\mathrm{a}}_{22}}
\newcommand{\omegaP}{\omega^{\mathrm{p}}_{22}}

\newcommand{\omegatwotwo}{\omega_{22}}
\newcommand{\omegaOrb}{\Omega_{\mathrm{orb}}}
\newcommand{\dotOmegaOrb}{\dot{\Omega}_{\mathrm{orb}}}

\newcommand{\eEOB}{e_{\mathrm{eob}}}
\newcommand{\eSpEC}{e_{\mathrm{SpEC}}\xspace}
\newcommand{\eSEOB}{e_{\mathrm{SEOB}}^{\mathrm{internal}}}
\newcommand{\lEOB}{l_{\mathrm{eob}}}

\newcommand{\freqref}{f_{\mathrm{ref}}}

\newcommand{\mResAmp}{\texttt{ResidualAmplitude}\xspace}

\newcommand{\mFreqFits}{\texttt{FrequencyFits}\xspace}
\newcommand{\mAmpFits}{\texttt{AmplitudeFits}\xspace}
\newcommand{\mSpline}{\texttt{spline}\xspace}
\newcommand{\mRatFit}{\texttt{rational\_fit}\xspace}

\newcommand{\tStart}{t_{0}}
\newcommand{\tStartHat}{\widehat{t}_{0}}
\newcommand{\tRefSpEC}{t^{\mathrm{ref}}_{\mathrm{SpEC}}}

\newcommand{\SEOB}{\texttt{SEOBNRv5EHM}\xspace}

\newcommand{\TEOB}{\texttt{TEOBResumS-Dal{\'i}}\xspace}

\newcommand{\SXS}{\texttt{SXS}\xspace}
\newcommand{\sxsID}[1]{\texttt{SXS:BBH:#1}}

\newcommand{\SpEC}{\texttt{SpEC}\xspace}

\newcommand{\package}{\texttt{gw\_eccentricity}~\cite{Shaikh:2023ypz}\xspace}

\newcommand{\tP}{t^{\mathrm{p}}}
\newcommand{\tA}{t^{\mathrm{a}}}
\newcommand{\copr}{\mathrm{copr}}
\newcommand{\ampCopr}{A^{\copr}}
\newcommand{\omegaCopr}{\omega^{\copr}}
\newcommand{\phaseCopr}{\phi^{\copr}}
\newcommand{\ampGW}{A_{\mathrm{gw}}}

\newcommand{\omegaGW}{\omega_{\mathrm{gw}}}
\newcommand{\omegaGWAS}{\omega^{\mathrm{as}}_{\mathrm{gw}}}
\newcommand{\omegaGWFiltered}{\omega^{\mathrm{filtered}}_{\mathrm{gw}}}
\newcommand{\phaseGW}{\phi_{\mathrm{gw}}}

\newcommand{\omegaGWP}{\omegaGW^{\mathrm{p}}}
\newcommand{\omegaGWA}{\omegaGW^{\mathrm{a}}}

\newcommand{\omegaAsymmetry}{\omega_{+}}
\newcommand{\eomegaGW}{e_{\omegaGW}}
\newcommand{\masym}{mode asymmetry\xspace}
\newcommand{\orbt}{orbital-timescale\xspace}
\newcommand{\dm}{(2, 2)\xspace}

\newcommand{\vL}{\boldsymbol{L}}

\newcommand{\vchi}{\boldsymbol{\chi}}
\newcommand{\sprec}{spin-precession\xspace}
\newcommand{\sprecing}{spin-precessing\xspace}
\newcommand{\aspin}{aligned-spin\xspace}
\newcommand{\aspins}{aligned-spins\xspace}

\newcommand{\sxsDemoId}{\texttt{SXS:BBH:3973}\xspace}

\newcommand{\deltaOmegaSpin}{\delta \omegaGW^{\mathrm{spin}}}
\newcommand{\deltaOmegaEcc}{\delta \omegaGW^{\mathrm{ecc}}}
\newcommand{\omegaGWSecular}{\omegaGW^{\mathrm{secular}}}
\newcommand{\fSpin}{f_{\mathrm{spin}}}
\newcommand{\fSpinIntersect}{f^{\mathrm{intersect}}_{\mathrm{spin}}}
\newcommand{\fCutoff}{f_{\mathrm{cutoff}}}
\newcommand{\fEcc}{f_{\mathrm{ecc}}}

\newcommand{\dtIntersect}{\Delta T^{\mathrm{intersect}}}
\newcommand{\rFilter}{r_{\mathrm{filter}}}

\usepackage{tikz}
\usetikzlibrary{shapes.geometric, arrows.meta, positioning}

\tikzstyle{startstop} = [rectangle, rounded corners, minimum width=3cm, minimum height=1cm,text centered, draw=black, fill=gray!10]
\tikzstyle{process} = [rectangle, minimum width=3.5cm, minimum height=1cm, text centered, draw=black, fill=blue!10]
\tikzstyle{decision} = [diamond, aspect=2, draw=black, fill=orange!20, text centered, inner sep=1pt]
\tikzstyle{arrow} = [thick, ->, >=stealth]

\newcommand{\AEI}{{Max Planck Institute for Gravitational Physics
    (Albert Einstein Institute), D-14476 Potsdam, Germany}}
\newcommand{\UMassD}{{Department of Mathematics,
    Center for Scientific Computing and Data Science Research,
    University of Massachusetts, Dartmouth, MA 02747, USA}}
\newcommand{\SNU}{{Department of Physics and Astronomy,
    Seoul National University, Seoul 08826, Korea}}
\newcommand{\VSM}{{Department of Physics, Vivekananda Satavarshiki Mahavidyalaya (affiliated to Vidyasagar University), Manikpara 721513, West Bengal, India}}
\newcommand{\UIB}{{Departament de F\'isica, Universitat de les Illes Balears, IAC3 -- IEEC, Crta. Valldemossa km 7.5, E-07122 Palma, Spain}}
\newcommand{\Cornel}{Cornell Center for Astrophysics and Planetary Science, Cornell University, Ithaca, New York 14853, USA}
\newcommand{\Caltech}{Theoretical Astrophysics 350-17, California Institute of Technology, Pasadena, CA 91125, USA}

\begin{document}

\title{Defining eccentricity for spin-precessing binaries}

\author{
  Md Arif Shaikh\,\orcidlink{0000-0003-0826-6164}$^{1,2}$\footnotemark[0],
  Vijay Varma\,\orcidlink{0000-0002-9994-1761}$^{3}$,
  Antoni Ramos-Buades\,\orcidlink{0000-0002-6874-7421}$^{4}$,
  Harald P. Pfeiffer\,\orcidlink{0000-0001-9288-519X}$^{5}$,
  Michael Boyle\,\orcidlink{0000-0002-5075-5116}$^{6}$,
  Lawrence E. Kidder\,\orcidlink{0000-0001-5392-7342}$^{6}$, and
  Mark A. Scheel\,\orcidlink{0000-0001-6656-9134}$^{7}$
}

\footnotetext[0]{Author to whom any correspondence should be addressed.}

\address{$^1$ \VSM}
\address{$^2$ \SNU}
\address{$^3$ \UMassD}
\address{$^4$ \UIB}
\address{$^5$ \AEI}
\address{$^6$ \Cornel}
\address{$^7$ \Caltech}

\ead{\mailto{arifshaikh.astro@gmail.com}}
\vspace{10pt}

\begin{abstract}
Standardizing the definition of eccentricity is necessary for
unambiguous inference of the orbital eccentricity of compact binaries
from gravitational wave observations. In previous works, we proposed a
definition of eccentricity for systems without spin-precession that
relies solely on the gravitational waveform, is applicable to any
waveform model, and has the correct Newtonian limit. In this work, we
extend this definition to spin-precessing systems. This simple yet
effective extension relies on first transforming the waveform from the
inertial frame to the coprecessing frame, and then adopting an
amplitude and a phase with reduced spin-induced effects. Our method
includes a robust procedure for filtering out spin-induced
modulations, which become non-negligible in the small eccentricity and
large spin-precession regime. Finally, we apply our method to a set of
Numerical Relativity and Effective One Body waveforms to showcase its
robustness for generic eccentric spin-precessing binaries. We make our
method public via Python implementation in \texttt{gw\_eccentricity}.
\end{abstract}

\vspace{2pc}
\noindent{\it Keywords}: Gravitational Waves, Compact Binary
Coalescences, Eccentricity, Numerical Relativity, Black Holes

\hypersetup{pdfauthor={Shaikh et al.}}

\date{\today}

\maketitle

\section{Introduction}\label{sec:introduction}
The ground-based network of gravitational wave (GW) detectors,
LIGO-Virgo-KAGRA
(LVK)~\cite{LIGOScientific:2014pky,TheVirgo:2014hva,KAGRA:2020tym},
has observed $\roughly$ 90 compact binary coalescences (CBCs)
during the first three observation runs (O1, O2, and
O3)~\cite{GWOSC:GWTC,GWOSC:GWTC-2,GWOSC:GWTC-2.1,GWOSC:GWTC-3}.
The increasing number of detected GW signals can enable us to address
one of the key scientific questions in GW astronomy—how do these
binaries form in nature? Two main formation channels are generally
considered for compact binaries: the first is the isolated formation
channel, where the binary evolves without interactions with any third
object~\cite{Mapelli:2021for,Mandel:2018hfr}. The second is the
dynamical formation channel, where the binary can undergo frequent
interactions with other objects in a dense stellar environment, such
as a globular cluster~\cite{Samsing:2017xmd,Zevin:2018kzq} or near an
active galactic
nucleus~\cite{Antonini:2016gqe,Samsing:2020tda,Tagawa:2020jnc}.

Eccentricity and the spins of black holes {provide valuable clues about}
 the formation history of a binary. Due to the loss of
energy and angular momentum via GW emission, the orbital eccentricity
of a binary decays over time as the system
inspirals~\cite{Peters:1963ux,Peters:1964zz}. As a result, for
binaries evolving in isolation, the orbit becomes circularized by the
time their GWs enter the sensitivity band of ground-based GW detectors.
On the other hand, in dense stellar environments, dynamical
interactions can harden the binary, leading to a merger with
non-negligible
eccentricity~\cite{Samsing:2017xmd,Zevin:2018kzq}. Similarly,
eccentricity can be amplified due to binary-single interactions in an
active galactic nucleus (AGN)
disk~\cite{Antonini:2016gqe,Samsing:2020tda,Tagawa:2020jnc} or via the
Kozai-Lidov mechanism in the presence of a third object near the
binary~\cite{Naoz:2016tri,Antonini:2017ash,Randall:2017jop,Yu:2020iqj}.

Similar to eccentricity, the formation history also influences the
spins of black holes. In the isolated formation scenario, the spins of
black holes are aligned with respect to the orbital angular momentum
of the binary~\cite{Mapelli:2021for,Mandel:2018hfr}. On the other
hand, in a dynamical environment, random interactions can produce
black hole spins with arbitrary
orientations~\cite{Mapelli:2021for}. When the spins are tilted
relative to the orbital angular momentum, spin-spin and spin-orbit
interactions cause the orbital plane to precess around the total
angular momentum of the
system~\cite{Apostolatos:1994pre,Kidder:1995zr}.  This has a direct
imprint on the amplitude and frequency of the GW signal, and can
therefore be used to infer the black hole spins from GW observations
as long as gravitational waveform models accurately capture the
effects of \sprec.

Similarly, accurately modeling the effects of eccentricity is
necessary to reliably measure eccentricity from GW observations.
Inclusion of eccentricity in waveform models is essential not only to
avoid systematic bias in the parameters inferred from GWs data
\cite{Favata:2021vhw,Islam:2021mha,OShea:2021faf,Clarke:2022fma,Ramos-Buades:2023yhy,DuttaRoy:2024aew,Nee:2025zdy,Huez:2025npe},
but also to avoid false violation of General Relativity (GR) in tests
of GR using
GWs~\cite{Narayan:2023vhm,Shaikh:2024wyn,Gupta:2024gun}. Accurate
quasicircular waveform models that capture \sprec have been
developed~\cite{Pan:2013rra,Hannam:2013oca,Taracchini:2013rva,Babak:2016tgq,Khan:2018fmp,Boyle:2014ioa,Blackman:2017pcm,Blackman:2017dfb,Varma:2019csw,Pratten:2020ceb,Ossokine:2020kjp,Akcay:2020qrj,Gamba:2021ydi,Estelles:2020osj,Estelles:2020twz,Estelles:2021gvs,Hamilton:2021pkf,Ramos-Buades:2023ehm,Thompson:2023ase,Colleoni:2024knd}. Similarly,
accurate eccentric \aspin waveform models have also been
developed~\cite{Tanay:2016zog,Moore:2018kvz,Moore:2019xkm,Hinderer:2017jcs,Khalil:2021txt,Ramos-Buades:2021adz,Gamboa:2024imd,Gamboa:2024hli,Nagar:2024dzj,Liu:2021pkr,Liu:2023dgl,Yun:2021jnh,Shi:2024age,Islam:2021mha,Joshi:2022ocr,Paul:2024ujx,Planas:2025feq,Islam:2025llx}.
However, to reliably extract the eccentricity and spins of black holes
from GWs, accurate waveform models that incorporate both eccentricity and
\sprec effects are required.
 
Ongoing efforts aim to develop waveform models that include both
effects, for example, using Numerical Relativity (NR)
simulations~\cite{Gayathri:2020coq,NRpaper}, Post-Newtonian (PN)
{theory}~\cite{Ireland:2019tao,Klein:2021jtd,Arredondo:2024nsl,Morras:2025nlp,Phukon:2025yva} and the effective one-body (EOB)
formalism~\cite{Liu:2023ldr,Gamba:2024cvy,Albanesi:2025txj}.  Several
{works} have already placed constraints on eccentricity of the
observed GW events~\cite{Romero-Shaw:2020thy, Gayathri:2020coq,
Romero-Shaw:2021ual, Gupte:2024jfe, Planas:2025jny, Morras:2025xfu,
Planas:2025plq}, but they require further refinement using improved
waveform models that capture the full physics of eccentricity and
\sprec (see~\cite{Gamba:2025qfg} for a first step in this direction).

A key challenge in constraining eccentricity from GW observations is
that eccentricity is not uniquely defined in GR. In fact, different
waveform models use different internal definitions of eccentricity,
which can lead to ambiguity in the inferred orbital
eccentricity. Several works have made efforts to standardize the
definition of
eccentricity~\cite{Knee:2022hth,Ramos-Buades:2022lgf,Shaikh:2023ypz,Bonino:2024xrv,Boschini:2024scu,Islam:2025oiv,Chartier:2025hsw}.
References~\cite{Ramos-Buades:2022lgf,Shaikh:2023ypz} proposed a
standardized definition of eccentricity that relies only on the
gravitational waveform, can be applied to any waveform model, and has
the correct Newtonian limit. However, the definition proposed
in~\cite{Ramos-Buades:2022lgf,Shaikh:2023ypz} assumes that the black
hole spins are either aligned or anti-aligned with the orbital angular
momentum, that is, the orbital plane remains fixed. In this work, we
relax this assumption and extend the definition of eccentricity to
systems with \sprec, where the black hole spins may be tilted with
respect to the orbital angular momentum.

We generalize the standardized definition of eccentricity by using the
waveform in the coprecessing frame and using variables with reduced
spin-induced effects. We show that our generalized definition can be
robustly applied to waveforms from different origins and broad range
of eccentricities. The robustness of our eccentricity estimation is
achieved through two additional improvements: an {adaptive}
interpolation method based on rational function approximation, and {an
additional processing step} for removing residual spin-induced
oscillations in the amplitude and frequency using low-pass filtering.
Additionally, we briefly explore the influence of the
gravitational-wave memory effect on eccentricity measurements, and
demonstrate that subtracting the memory contribution leads to more
consistent and reliable eccentricity estimates.

The rest of the paper is organized as follows: In
section~\ref{sec:generalizing_eccentricity_definition}, we discuss the
methods for generalizing the definition of eccentricity, and
provide a summary of the steps required to measure eccentricity for
\sprecing systems. In section~\ref{sec:results}, we apply the
generalized definition across different waveform models and
eccentricities to check the robustness of our method, and
conclude in section~\ref{sec:conclusion}.

\section{Generalizing the definition of eccentricity to spin-precessing binaries}
\label{sec:generalizing_eccentricity_definition}

\subsection{Notation and conventions}
\label{sec:waveform_modes_and_conventions}
GW emitted by a compact binary can be represented as a complex
combination $\h$ of its two polarizations, \( h_{+} \) (plus) and \(
h_{\times} \) (cross): $\h \equiv h_{+} - i h_{\times}$. The complex
waveform $\h$ can be decomposed into a sum of spin-weighted spherical
harmonic modes $\hlm$ such that the waveform along any direction
$(\iota, \varphi_0)$ in the binary's source frame can be expressed as
\begin{equation}
\label{eq:spherical-harmonics}
\h(t, \iota,\varphi_0)
    = \sum_{\ell=2}^{\ell=\infty}\sum_{m=-\ell}^{m=\ell}
    \hlm(t) ~ \ylm(\iota, \varphi_0),
\end{equation}
where $\iota$ and $\varphi_0$ are the polar and the azimuthal angles,
respectively, on the sky in the source frame of the binary. $\ylm$ are the
spin=-2 weighted spherical harmonics.

For a given complex mode $\h_{\ell,m}$, we can define corresponding
amplitude $\amplm$, phase $\philm$ and the angular
frequency $\omega_{\ell, m}$ as
\begin{eqnarray}
  \h_{\ell,m} = \amplm ~ e^{-i\philm},\label{eq:hlm_in_amp_phase}\\
  \omega_{\ell, m} = \frac{\dd \philm}{\dd t}\label{eq:omegalm}.
\end{eqnarray}
For a binary system on a generic bound orbit, the waveform modes
$\hlm(t)$ depend on the component masses ($m_1$ and $m_2$, with \( m_1
\geq m_2 \)), six spin components (three for each compact object), two
eccentricity parameters (eccentricity and mean anomaly), and the
luminosity distance. The eccentricity parameters and black hole spins
(for \sprecing binaries) evolve over time, and, therefore these
quantities should be defined at a fixed reference time (or frequency).
Unless explicitly specified, we work with $\hlm(t)$ at future null
infinity scaled to unit total mass and luminosity distance. Thus, the
system at a given reference time (or frequency) is characterized by 9
parameters --- the mass ratio ($ q = m_1 / m_2 $), the dimensionless
spin vectors ($ \vchi_1, \vchi_2 $), and the two eccentricity
parameters. Finally, it is convenient to define an effective
spin-precessing parameter $\chi_p$ that approximately quantifies the
amount of \sprec in a system~\cite{Schmidt:2014iyl}
\begin{equation}
  \label{eq:chi_p_definition}
  \chi_p = \mathrm{max}\left(\chi_1\sin\theta_1, \frac{4q + 3}{4 + 3q} q
  \chi_2\sin\theta_2\right),
\end{equation}
where $\chi_i$ is the component spin magnitude with $i=1,2$ and
$\theta_i$ is the component spin tilt relative to the orbital angular
momentum.

Throughout the paper, we shift the time array of the waveform such that $t=0$
occurs at the peak of the waveform amplitude defined in equation (5)
of~\cite{Varma:2019csw}~\footnote{While $t = 0$ according to this definition
coincides with the merger time for most systems, this may not always hold for
highly eccentric systems. To ensure that $t = 0$ corresponds to the merger
time, one can define $t = 0$ as the time of the last peak in the waveform
amplitude {(see discussion near equation (14)
of}~\cite{Gamboa:2024hli}).}. Eccentricity, mean anomaly, and black hole spins
(for \sprecing binaries) are presented at a fixed time before the peak.

\subsection{Effect of eccentricity and spin-precession {on the waveform}}\label{sec:effect_of_eccentricity_and_spin_precession}
\begin{figure}[t]
  \centering
  \includegraphics{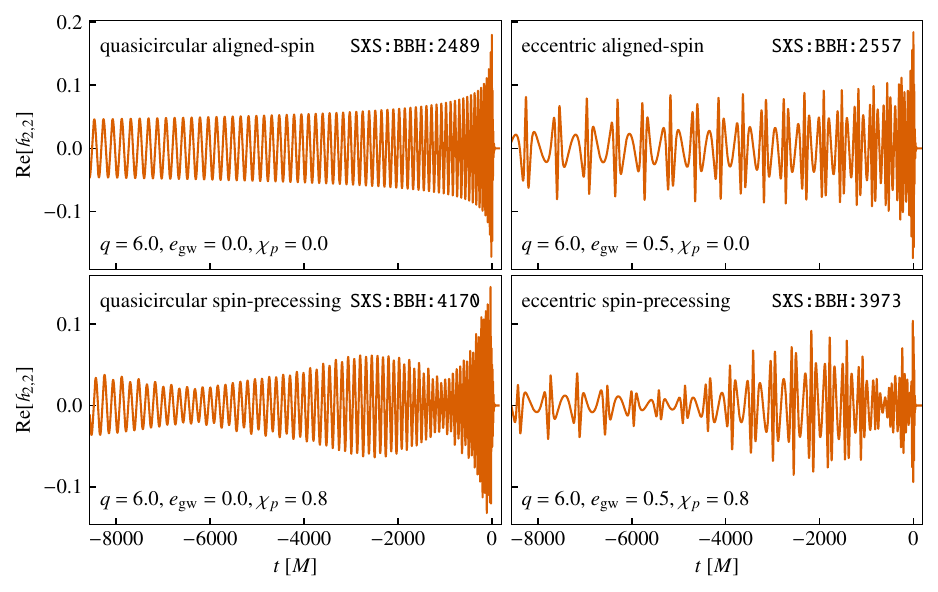}
  \caption{{\bfseries Effects of eccentricity and \sprec on waveforms}. 
    {\itshape Top-left:} A quasicircular \aspin system, where we can
    see the monotonic increase of the amplitude and frequency. {\itshape
      Top-right:} An eccentric \aspin system, where modulations due to
    eccentricity are visible on the orbital timescale. {\itshape
      Bottom-left:} A quasicircular \sprecing system, where modulations due
    to \sprec are seen over the longer \sprec timescale (spanning several
    orbits). {\itshape Bottom-right:} An eccentric \sprecing
    system. Features of eccentricity on \orbt and features of \sprec on
    spin-precession timescale are simultaneously visible. In all panels,
    the eccentricity $\egw$ is measured at $t = -8000M$, following the
    procedure described in
    section~\ref{sec:waveform_modes_and_conventions}.  The inset texts
    indicate the simulation IDs from the SXS
    Catalog~\cite{Scheel:2025jct}. We have removed the memory contribution
    from these waveforms, which is included by default in the latest
    catalog of SXS waveforms.
        }
  \label{fig:effects_of_eccentricity_and_precession}
\end{figure}
The GW modes {$\hlm$} exhibit the simplest morphology when the binary
system follows a quasicircular orbit with no \sprec. The GW emitted by
such a system displays a monotonically increasing behavior in both
amplitude and frequency. In
figure~\ref{fig:effects_of_eccentricity_and_precession}, the {\itshape
top-left} plot shows the real part of $\h_{2,2}$ for a quasicircular,
\aspin (nonprecessing) system simulated using the Spectral Einstein
Code (\SpEC)~\cite{SpECwebsite,SXSCatalog,Scheel:2025jct} developed
by the Simulating eXtreme Spacetime (\SXS)
collaboration~\cite{SXSWebsite}. We work with waveforms extrapolated
to future null infinity, with extrapolation order $N=2$. The latest
catalog of SXS simulations includes memory by
default~\cite{Scheel:2025jct,Mitman:2020bjf}. The memory correction
is performed by adding the memory term to the extrapolated waveforms,
which is computed using equation (11) in \cite{Scheel:2025jct}.
Unless stated otherwise, we undo this inclusion and work with memory
removed waveforms (as will be expanded upon in
section~\ref{sec:effects_of_waveform_memory}).

By contrast, orbital eccentricity results in bursts of GW radiation at
every pericenter (point of closest approach) passage
\cite{Peters:1963ux,Peters:1964zz}, which appear as modulations of the
GW amplitude and frequency on the \orbt. The {\itshape
top-right} plot in
figure~\ref{fig:effects_of_eccentricity_and_precession} shows the real
part of $\h_{2,2}$ for an \aspin system on an eccentric orbit.

In an \aspin system, the spins of the black holes are either aligned or
anti-aligned with respect to the orbital angular momentum $\vL$ which remains
fixed throughout the system's evolution. The $\hat{z}$ direction of the
binary's source frame is choosen to be along $\vL$ {by} convention. In such
systems, the direction of the strongest GWs emission is
parallel and anti-parallel to $\vL$. Consequently, the ($\ell=2, m=\pm 2$)
modes remain the dominant modes during the binary's entire evolution. In
addition, due to the symmetry about the orbital plane (which lies on the $x-y$
plane), the negative and positive $m$ modes are related by
\begin{equation}
\label{eq:mode_symmetry_aligned-spin}
\h_{\ell, m} = (-1)^\ell \h^{*}_{\ell,-m},
\end{equation}
where $*$ stands for complex conjugation.
On the other hand, in a \sprecing system, the black hole spins are
tilted relative to the orbital angular momentum $\vL$, causing the
orbital angular momentum and the spins to precess around the total
angular momentum of the system due to spin-spin and spin-orbit
interactions~\cite{Apostolatos:1994pre}. Therefore, the $\hat{z}$
direction of the source frame, which by convention is aligned with
$\vL$ at a given reference time or frequency, no longer coincides with
$\vL$ as the binary inspirals. This leads to leakage of GW power into
($\ell=2, m\neq \pm 2$) modes from the ($\ell=2, m=\pm 2$) modes
causing modulation of the waveform on the {longer} \sprec timescale{,
which occurs over several orbits}. In
figure~\ref{fig:effects_of_eccentricity_and_precession}, the {\itshape
bottom-left} plot shows the real part of $\h_{2,2}$ for a
quasicircular \sprecing system.

Finally, when the binary follows the most generic bound orbit,
incorporating both \sprec and eccentricity, the waveform modes exhibit
the most complex morphology, showing the imprints of both effects. In
figure~\ref{fig:effects_of_eccentricity_and_precession}, the {\itshape
bottom-right} plot shows the real part of $\h_{2,2}$ of an eccentric
\sprecing binary. The waveform shows modulations over orbital as well
as \sprec timescale.

\subsection{Challenges in defining eccentricity in presence of spin-precession}\label{sec:challenges_in_defining_eccentricity_in_presence_of_spin-precession}
Due to the leakage of GW power from the $(\ell=2,m=\pm 2)$ modes to
($\ell=2,m \neq \pm 2$) modes in the inertial frame for a \sprecing
system, the \dm mode is not necessarily the dominant mode throughout the binary’s
evolution, and other ($\ell=2, m\neq \pm2$) modes may become comparable
at different times during the evolution (see e.g. figure 1 of
~\cite{Varma:2019csw}).

For an \aspin system on an eccentric orbit, the frequency
$\omegatwotwo$ of the inertial frame \dm mode exhibits the property
that, while $\omegatwotwo$ itself is nonmonotonic, $\omegaP$ (the
values of $\omegatwotwo$ at pericenters) and $\omegaA$ (the values of
$\omegatwotwo$ at apocenters) are monotonically increasing functions
of time (see e.g. figure 1 of~\cite{Shaikh:2023ypz}).  This property
enables the construction of monotonic interpolants, $\omegaP(t)$ (the
interpolant {through} $\omegatwotwo$ at pericenters) and $\omegaA(t)$ (the
interpolant {through} $\omegatwotwo$ at apocenters). These monotonic
interpolants were crucial for obtaining a consistent monotonic
evolution of eccentricity in~\cite{Shaikh:2023ypz}. Thus, for
\aspin systems, it suffices to work with the inertial frame
\dm mode for extracting eccentricity information from its frequency
evolution~\cite{Shaikh:2023ypz}.

However, in \sprecing systems, $\omegaP$ and $\omegaA$ computed in the
inertial frame do not follow a monotonic trend. Therefore, the
inertial-frame \dm mode frequency $\omegatwotwo$, on its own, is not
suitable for measuring eccentricity in \sprecing systems.

\subsubsection{Coprecessing frame}
\label{sec:coprecessing_frame}
To address the challenges of defining eccentricity for a \sprecing
system, we aim to minimize the effects of \sprec in the waveforms used
for computing eccentricity. The primary challenge lies in leakage of
GW power into $(\ell=2, m \neq \pm 2)$ modes from $(\ell=2, m = \pm
2)$ modes, which arises because the angular momentum $\vL$ evolves
over time, while the frame used to decompose the waveform into
spin-weighted spherical harmonics remains static. A straightforward
solution to mitigate this is to transition to the coprecessing frame,
which is a non-inertial, time-dependent frame where the
$\hat{z}$-axis of the frame aligns with the direction of maximal
gravitational wave emission (which closely follows $\vL$) at every
instant~\cite{Schmidt:2010it,OShaughnessy:2011pmr,Boyle:2011gg}. The
waveform modes $\hlm$ are then expressed relative to this dynamic
basis, thereby reducing the impact of precession of $\vL$ on the mode
decomposition.  This strategy has been used to simplify the modeling
of \sprecing waveforms in the quasicircular case
~\cite{Estelles:2020osj,Estelles:2020twz,Estelles:2021gvs,Akcay:2020qrj,Gamba:2021ydi,Hamilton:2021pkf,Ramos-Buades:2023ehm}
{(and more recently in the eccentric case
in~\cite{Gamba:2024cvy})}. In this work, we utilize the frame rotation
functionality provided by the public library \texttt{scri}~\cite{scri}
to transform the waveform modes to the coprecessing frame. By design,
the coprecessing frame ensures that the $(\ell=2, m=\pm 2)$ modes
remain the dominant modes throughout the binary's evolution in this
frame. For example, see the {\itshape middle} panel of figure 1 in
\cite{Varma:2019csw}.

\begin{figure}
  \centering
  \includegraphics{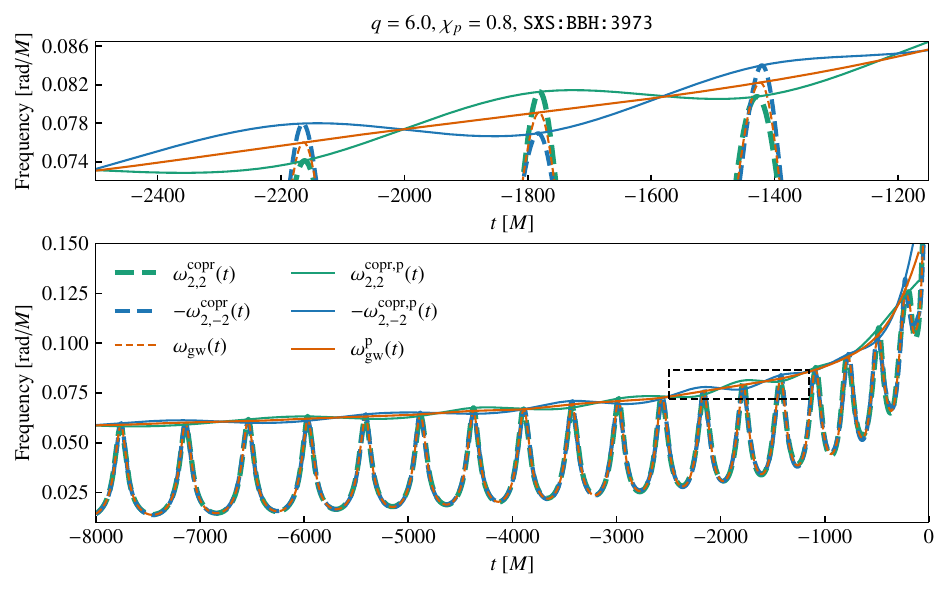}
  \caption{{\bfseries GW frequency in the coprecessing frame.}
{\itshape Bottom:} Shows the coprecessing frame frequencies
$\omegaCopr_{2, 2}$, $\omegaCopr_{2, -2}$, and their antisymmetric
combination $\omegaGW$ {as dashed curves of different thickess}. It also shows the interpolants through their
values at their respective local maxima (denoted by a
superscript ``p" for {\itshape pericenters}) as solid curves. Both
$\omegaCopr_{2, 2}$ and $\omegaCopr_{2,-2}$ exhibit a reduced
influence of \sprec compared to their inertial frame counterparts, and
closely resemble the frequency evolution of an \aspin
system. However, they still retain \orbt spin-induced effects,
as highlighted in the {\itshape top} panel. 
{\itshape Top:} 
A zoomed-in view of the boxed region in the bottom panel.
Spin-precession affects $\omegaCopr_{2, 2}$
and $\omegaCopr_{2,-2}$ in nearly opposite ways on the \orbt,
leading to their alternate dominance. As a result, the values of these
frequencies at the pericenters becomes oscillatory, introducing
nonmonotonicity in their interpolants (solid blue and green). In
contrast, $\omegaGW$ displays a monotonic increase in its values at
the pericenters (solid orange). {The system has an eccentricity
  $\egw = 0.5$ at $t=-8000M$.}
} 
  \label{fig:omegas_in_coprecessing_frame}
\end{figure}

Although, the mode hierarchy is reestablished in the coprecessing
frame, it cannot get rid of all {\sprec} effects. In particular, for
\sprecing systems, there exists no frame where the
equation~(\ref{eq:mode_symmetry_aligned-spin}) is
true~\cite{Boyle:2014ioa}, {leading to a {\itshape \masym} between the
positive and negative $m$ modes for a given $\ell$. In addition to the
multiple-orbit-timescale motion of the orbital-plane described above,
\sprec induces waveform} features varying on the \orbt
that cannot be {removed} by applying any
rotation~\cite{Boyle:2014ioa}. Therefore, even in the coprecessing
frame, the frequency of the $\dm$ mode is not guaranteed to have a
monotonically increasing frequency at the pericenters and
apocenters. One can, however, get around this by noticing that, at any
given instant, the $(\ell, m)$ and $(\ell, -m)$ modes are affected
nearly oppositely by the
\sprec~\cite{Boyle:2014ioa,Blackman:2017pcm,Blackman:2017dfb,Varma:2019csw}.
Consequently, the $(2, 2)$ and $(2,-2)$ modes alternately dominate the
GW power on the \orbt, as illustrated in
figure~\ref{fig:omegas_in_coprecessing_frame} through their
frequencies. We denote the quantities in the coprecessing frame using
the superscript ``copr''.

To generalize the definition of eccentricity to \sprecing systems, one could
try to replace $\omegaP$ and $\omegaA$ in
section~\ref{sec:challenges_in_defining_eccentricity_in_presence_of_spin-precession}
with the values of $\omegaCopr_{2,2}$ and $\omegaCopr_{2,-2}$ evaluated at the
pericenters and apocenters. However, spin-induced \orbt effects break the
monotonicity of the values of $\omegaCopr_{2,2}(t)$ and $\omegaCopr_{2,-2}(t)$
at their extrema (interpolants through these quantities at the pericenters are
denoted as $\omega^{\mathrm{copr,p}}_{2,2}$ and
$\omega^{\mathrm{copr,p}}_{2,-2}$ in
figure~\ref{fig:omegas_in_coprecessing_frame} ). To mitigate this, we follow
~\cite{Boyle:2014ioa,Blackman:2017pcm,Blackman:2017dfb,Varma:2019csw}, and
define quantities in the coprecessing frame by combining the $(2, 2)$ and $(2,
-2)$ modes
\begin{eqnarray}
\label{eq:amp_gw} \ampGW = \frac{1}{2}\left(\ampCopr_{2,2} +
\ampCopr_{2,-2}\right),\\
\label{eq:phase_gw} \phaseGW = \frac{1}{2}\left(\phaseCopr_{2,2} -
\phaseCopr_{2,-2}\right),\\
\label{eq:omega_gw} \omegaGW = \frac{\dd \phaseGW}{\dd t} =
\frac{1}{2}\left(\omegaCopr_{2, 2} - \omegaCopr_{2, -2}\right).
\end{eqnarray} In literature, these quantities have previously been denoted as
$A_{+}, \phi_{-}$ and $\omega_{-}$, respectively (see e.g. equations (48) and
(49) in ~\cite{Blackman:2017dfb}). {Furthermore, in the quasicircular
\sprecing case, the integral of $\omegaGW$ over time serves as a proxy for the
orbital phase~\cite{Varma:2019csw}.} We use the notation in equations
(\ref{eq:amp_gw})-(\ref{eq:omega_gw}) for convenience in the equations that
appear in the rest of the paper.

In case of an \aspin system, quantities in equations
(\ref{eq:amp_gw})-(\ref{eq:omega_gw}) reduce to the corresponding
values of the (2, 2) mode because the inertial and coprecessing frames
are equivalent in \aspin systems, and the negative and positive
$m$ modes are related via
equation~(\ref{eq:mode_symmetry_aligned-spin}) due to the symmetry
about the orbital plane.

The {\itshape bottom} panel of
figure~\ref{fig:omegas_in_coprecessing_frame} shows the coprecessing
frame frequencies $\omegaCopr_{2,2}$ and $\omegaCopr_{2,-2}$ and
their antisymmetric combination $\omegaGW$. These frequencies
correspond to an NR waveform of an eccentric \sprecing system
\sxsDemoId with $q=6$ and $\chi_p=0.8$.
The panel also shows the interpolants (denoted by a superscript ``p'') of
their values at the pericenters. The {\itshape top} panel focuses on
a region near the pericenters (boxed region in the {\itshape bottom}
panel). While $\omega^{\mathrm{copr,p}}_{2,2}$ and
$\omega^{\mathrm{copr,p}}_{2,-2}$ at the extrema dominate alternately,
the {values at extrema} for $\omegaGWP$ increase monotonically.

As an alternative to $\omegaGW$, one can also define a rotationally
covariant angular velocity, defined as the velocity of the rotating
frame that minimizes the time dependence of the
modes~\cite{Boyle:2013nka}. However, since we find no significant
difference between this angular velocity and $\omegaGW$, we choose to
use $\omegaGW$ to define $\egw$.

\subsection{Generalized eccentricity definition}\label{sec:generalized_eccentricity_definition}
We extend the definition of eccentricity to \sprecing systems using
the GW frequency $\omegaGW$ defined in
equation~(\ref{eq:omega_gw}). The steps for computing $\egw$ and
$\lgw$ are identical to those outlined in section II.H of
~\cite{Shaikh:2023ypz}, with one key distinction: {we use} the
variables $\ampGW$, $\phaseGW$, and $\omegaGW$ (defined in
Eqs.~(\ref{eq:amp_gw}), equation~(\ref{eq:phase_gw}), and
equation~(\ref{eq:omega_gw}), respectively) instead of the
corresponding quantities from the \dm mode in the inertial
frame. First we define $\eomegaGW$ using $\omegaGW$,
\begin{equation}
  \eomegaGW(t) = \frac{\sqrt{\omegaGWP(t)} - \sqrt{\omegaGWA(t)}}
  {\sqrt{\omegaGWP(t)} + \sqrt{\omegaGWA(t)}}\label{eq:eomegagw}.
\end{equation}
Here, $\omegaGWP(t)$ and $\omegaGWA(t)$ are the interpolants through the
$\omegaGW$ values at the pericenters $(\omegaGWP)$ and apocenters
$(\omegaGWA)$, respectively.
We then define the eccentricity $\egw$
using the following transformation to ensure that it correctly
exhibits the Newtonian limit~\cite{Ramos-Buades:2022lgf}.

\begin{equation}
  \label{eq:egw}
  \egw(t) = \cos\left(\frac{\Psi(t)}{3}\right) - \sqrt{3}\sin\left(\frac{\Psi(t)}{3}\right),
\end{equation}
where
\begin{equation}
  \label{eq:Psi}
  \Psi(t) = \arctan\left( \frac{1 - \eomegaGW^2(t)}{2\eomegaGW(t)} \right).
\end{equation}

Once we obtain the pericenter times $\tP$, the mean anomaly between
two successive pericenters, $\tP_{i}$ and $\tP_{i+1}$, can be defined
over the interval $\tP_{i} \leq t < \tP_{i+1}$ as
\begin{equation}
  \label{eq:mean_anomaly}
  \lgw(t) = 2\pi \frac{t - \tP_{i}}{\tP_{i+1} - \tP_{i}}.
\end{equation}

To construct the interpolants $\omegaGWP(t)$ and $\omegaGWA(t)$, we
require the values of $\omegaGW$ at the pericenters and apocenters,
respectively. These are identified as the local maxima and minima of
either $\omegaGW$ or $\ampGW$. However, in systems with small
eccentricity, these extrema can be difficult to detect directly due to
the eccentric modulations becoming small compared to the secular
growth in $\omegaGW$ or $\ampGW$.  To address this, we work with the
residuals obtained by subtracting the secular trend from $\omegaGW$ or
$\ampGW$, thereby enhancing the prominence of the oscillatory
features. The secular trend can be estimated either by fitting a
power-law model inspired by PN expressions in the quasicircular limit
(this choice using $\ampGW$ is referred to as
$\mAmpFits$~\cite{Shaikh:2023ypz}), or by using the amplitude or
frequency of a corresponding quasicircular waveform (this choice using
$\ampGW$ is referred to as $\mResAmp$~\cite{Shaikh:2023ypz}). The
quasicircular waveform is generated using the same binary parameters
as the eccentric waveform, with the eccentricity set to zero. A
detailed discussion of various methods for locating extrema using
frequency- and amplitude-derived quantities can be found in Section
III of \cite{Shaikh:2023ypz}. Since the locations of the extrema may
vary slightly depending on the choice of data, the extrema-finding
procedure should be regarded as a part of the eccentricity
definition. In this work, unless stated otherwise, we use the
$\mAmpFits$ method to locate the extrema for measuring eccentricity
and mean anomaly throughout the rest of the paper.

Finally, $\phaseGW$ is used for internal diganostics and for estimating
the orbital period when necessary.  One must also be cautious about
numerical noise and ensure that the identified extrema arise from
genuine eccentricity-induced modulations, rather than spurious
fluctuations. Since eccentricity leads to oscillations on the \orbt,
the phase difference in $\phaseGW$ between two consecutive extrema
should be approximately $4\pi$ (as $\omegaGW$ is roughly twice the
orbital frequency). We use this criterion to discard extrema that are
likely due to numerical noise.

\subsection{Effects of waveform memory}\label{sec:effects_of_waveform_memory}
\begin{figure}[t]
  \centering
  \includegraphics{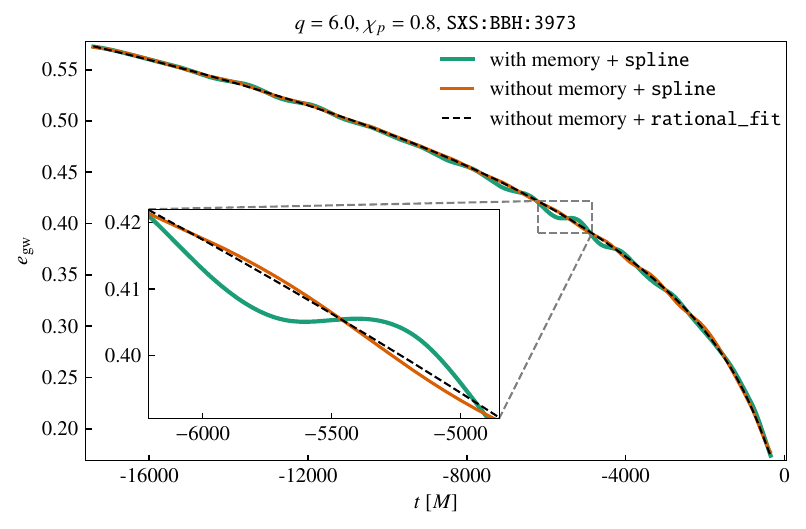}
  \caption{{\bfseries Effect of memory on $\egw(t)$}. For unequal mass
    and highly spin-precessing systems, the memory causes visible
    oscillations in the measured $\egw(t)$. The $\egw(t)$ after memory removal
    gets rid of most of these oscillations. Some oscillations remain
    in $\egw(t)$ even after memory removal, which go away 
    when we switch from \mSpline interpolation to \mRatFit
    when constructing the $\omegaP(t)$ and $\omegaA(t)$ interpolants
    in equation~(\ref{eq:eomegagw})
    (discussed in detail in
    section~\ref{sec:rational_fits_for_omega_extrema}). The $\egw(t)$
    before and after removal of memory is shown for \sxsDemoId.
    \label{fig:memory_effect}}
\end{figure}
Our goal is to extract the orbital eccentricity from the GW
signal. However, unlike the orbital dynamics, the waveforms at future
null infinity can contain imprints of additional effects, such as
gravitational
memory~\cite{Zeldovich:1974gvh,Braginsky:1985vlg,Braginsky:1987kwo,Christodoulou:1991,Thorne:1992memory,Favata:2010zu,Mitman:2024uss}.
Therefore, $\egw$ computed using GW can inherit a memory dependence.
This can be attributed to the fact that memory effect alters the
average amplitude of the waveform modes, which in turn, affects the
transformation of the waveform to the coprecessing frame and the
quantities such as $\omegaGWP(t)$ and $\omegaGWA(t)$ used to define
$\egw$. More specifically, the angular momentum operator, when applied
on the waveform modes, returns an extra term due to the memory
contribution (see the discussion around equation (11) and (12)
in~\cite{Boyle:2013nka}), which can impact the direction of maximal GW
emission used to define the coprecessing frame.

To avoid this, we perform memory removal from the waveforms before
using them to measure $\egw$. In figure~\ref{fig:memory_effect}, we
demonstrate the effect of memory on $\egw$ for an unequal mass ($q=6$)
and highly spin-precessing ($\chi_p = 0.8$) NR waveform. Without
memory removal the $\egw$ curve has small oscillations which are
reduced when we use waveforms after memory removal. Some modulations
remain in $\egw(t)$ even after memory removal, which go away with
\mRatFit interpolation for $\omegaGWP(t)$ and $\omegaGWA(t)$ , which
will be discussed in the next section. Throughout the rest of the
paper, we will use waveforms after memory removal.

{As mentioned earlier, we work with the waveforms extrapolated to the
future null infinity, with extrapolation order $N=2$. Waveforms with a different extrapolation order may differ slightly from the $N=2$
waveforms.
We explore the dependence of $\egw$ on the extrapolation order in
\ref{sec:dependence_on_waveform_extraction_method} for the highly
eccentric and \sprecing system \sxsDemoId, and find that
the differences in $\egw$ due to different extrapolation orders is comparable to
differences due to NR truncation error.
}

\subsection{Rational Fits for $\omegaGWP(t)$ and $\omegaGWA(t)$}\label{sec:rational_fits_for_omega_extrema}
\begin{figure}[h]
  \centering
  \includegraphics{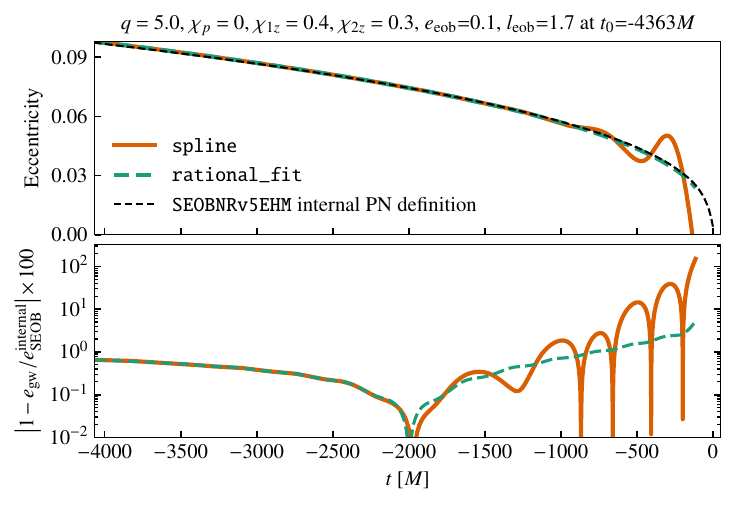}
  \caption{{\bfseries Comparison of Interpolation Methods}. {\itshape
Top:} $\egw(t)$ measured (using \mResAmp) from a waveform generated
using \SEOB, with the initial input parameters specified in the plot
title, where $\eEOB$ and $\lEOB$ are the initial input model
eccentricity and mean anomaly, respectively, at the starting time
$\tStart$. Using \mSpline for $\omegaGWP(t)$ and $\omegaGWA(t)$ leads
to nonmonotonic oscillations in the computed $\egw$ near the
merger. In contrast, when using the rational function approximation
(\mRatFit), the resulting $\egw$ is free of these oscillations. For
reference, we also include the eccentricity $\eSEOB$, the internal
definition of eccentricity in the Keplerian parametrization of the
conservative dynamics in \SEOB. {\itshape Bottom:} the instantaneous
percentage difference between $\egw$ using \mSpline or \mRatFit and
$\eSEOB$.}
  \label{fig:interpolation_method_comparison_demo}
\end{figure}
The definition of eccentricity $\egw$ in equation~(\ref{eq:egw}) relies on
the interpolants $\omegaP(t)$ and $\omegaA(t)$. In our previous work
in ~\cite{Shaikh:2023ypz}, we used Cubic interpolating splines\footnote{Uses
\texttt{scipy.interpolate.InterpolatedUnivariateSpline}} to build
these interpolants. We find that while such interpolants work very well for most
part of the inspiral, they can struggle at times close to the
merger. Near the merger, the \mSpline interpolants introduces 
artificial oscillations in the measured $\egw$ (see
figure~\ref{fig:interpolation_method_comparison_demo}). To avoid such 
oscillations in $\egw$ near the merger, we seek alternative
interpolation methods that are robust near the merger.

Following previous works~\cite{Bonino:2024xrv}, we employ an
alternative interpolation method for modeling 
$\omegaGWP(t)$ and $\omegaGWA(t)$ using rational
functions. While~\cite{Bonino:2024xrv} uses rational function
  of the form $(a + bx)/(1 + cx)$, we find it to be {less effective} for
  fitting $\omegaGWP(t)$ and $\omegaGWA(t)$ for longer
  waveforms. Therefore, we build the interpolants using rational
  functions with higher numerator and denominator degrees when
  required. Specifically, we employ the ``Stabilized
Sanathanan-Koerner'' iterations to build rational
approximations~\cite{SKA}\footnote{We use the \texttt{Python}
  implementation \texttt{polyrat.StabilizedSKRationalApproximation} of 
  the algorithm~\cite{polyrat,SKA}}. We refer to this method simply as \mRatFit.

To ensure that the $\omegaGWP(t)$ and $\omegaGWA(t)$ interpolants
are both monotonically increasing and accurate, we carefully select an
optimal pair of numerator and denominator degrees for the rational
function approximation algorithm. If the degrees are too high, spurious
divergences may appear in the interpolants; if too low, the resulting
interpolants may be insufficiently accurate. To address this, we
adaptively adjust the numerator and denominator degrees used in the
rational function approximation algorithm.

We begin with an initial guess for the degrees based on the
approximate number of orbital cycles (which can be estimated from the
maxima) in the waveform. We then compute the first
time derivative of the resulting interpolant. If this derivative is
strictly monotonic, we increment both degrees by one to improve
accuracy and repeat the process. If the derivative is not strictly
monotonic, the chosen degrees are too high, and we reduce them by one
until monotonicity is restored or the degrees reach a minimum value
of one. This iterative procedure eliminates divergences in the
interpolants and ensures a smooth, monotonically increasing evolution
of $\egw$ when using \mRatFit.

We find that \mRatFit performs as robustly and accurately as \mSpline
in regions far from the merger while showing superior robustness and
accuracy near the merger. This is evident in
figure~\ref{fig:interpolation_method_comparison_demo}, where the
{\itshape top} panel compares $\egw$ obtained using \mSpline and
\mRatFit to $\eSEOB$, the internal definition of eccentricity in
the Keplerian parametrization of the conservative dynamics in
\SEOB~\cite{Gamboa:2024hli,Gamboa:2024imd}. The {\itshape bottom}
panels shows the instantaneous percentage
difference. Figure~\ref{fig:interpolation_method_comparison_demo}
demonstrates that \mRatFit provides a measurement of $\egw$ that is
more consistent with $\eSEOB$, while being free of the spurious
oscillations present in \mSpline near the
merger. In~\ref{sec:rational_fit}, we compare these two methods on a
larger set of waveforms to assess their robustness.  We find that the
differences between $\eSEOB$ and $\egw$ can reach $\sim 1000\%$ when
using \mSpline, especially for small eccentricities ($\egw \lesssim 5
\times 10^{-2}$) and near the merger. For the same cases, the
differences between $\eSEOB$ and $\egw$ fall within $10\%$ with
\mRatFit.

In addition to providing robust interpolation near the merger,
\mRatFit { is also less susceptible to} small oscillations {in the
inspiral that may arise} due to numerical noise, and provides smoother
monotonically decreasing $\egw$ compared to \mSpline. {When} building
the $\omegaGWP(t)$ and $\omegaGWA(t)$ interpolants, \mRatFit minimizes
the $L_2$ norm of the error between the data and the approximation
{rather than strictly interpolate through the data points}.  Thus
it approximates the overall trend of the data which helps avoid local
fluctuations present when using \mSpline.  Nevertheless, one must be
cautious when using the GW frequency obtained from noisy NR waveforms,
as it can lead to spurious oscillations in $\egw$. In
\ref{sec:eccentricity_measurement_in_presence_of_numerical_noise} we
discuss how the oscillations in $\egw$ computed with \package noted
by~\cite{Islam:2025oiv} arise mainly due to numerical noise in the 
NR waveforms, and how \mRatFit is less susceptible to such
oscillations.

Similar to numerical noise, memory effects can
also introduce oscillations in $\egw(t)$ as discussed in
section~\ref{sec:effects_of_waveform_memory}. While removing the
memory significantly suppresses these oscillations, some residual
oscillations remain in $\egw(t)$ when computed using \mSpline. These
residual oscillations disappear when using \mRatFit, as shown in
figure~\ref{fig:memory_effect}. Throughout the rest of the paper, we
will use \mRatFit when computing $\egw$.

\subsection{Small eccentricity \& large spin-precession regime}
\label{sec:low_eccentricity_and_high_precession}
By transforming the waveform to the coprecessing frame, we are able to
remove the effects of the precession of the orbital plane on the
waveform to a large extent. However, as discussed in
section~\ref{sec:coprecessing_frame}, the waveform in the coprecessing
frame still exhibits \masym caused by \sprec. The effect of \masym
on $\egw$ is reduced by using $\omegaGW$ instead of $\omegaCopr_{2,2}$
in defining eccentricity in equation~(\ref{eq:egw}). While this is
sufficient to obtain a monotonically decreasing $\egw(t)$ for large
eccentricity, we need to take further steps to ensure
correct $\egw(t)$ measurement in the small eccentricity and large
\sprec regime.

One can express $\omegaGW$ as the sum of a secular, monotonically
increasing trend ($\omegaGWSecular$) caused by
radiation reaction and modulations arising due to eccentricity
($\deltaOmegaEcc$) and spin-induced effects
($\deltaOmegaSpin$)~\cite{Buonanno:2010yk,Knapp:2024yww,Habib:2024soh}:
\begin{equation}
  \label{eq:omega_decomposition}
  \omegaGW = \omegaGWSecular + \deltaOmegaEcc +
  \deltaOmegaSpin.
\end{equation}
The amplitude of $\deltaOmegaSpin$, compared to $\deltaOmegaEcc$,
depends on the eccentricity $\egw$, the \sprec
parameter $\chi_p$, and the mass ratio $q$ of the system. For large
eccentricity, $\deltaOmegaSpin \ll \deltaOmegaEcc$. However, at
smaller eccentricities, the spin-induced oscillation $\deltaOmegaSpin$
can become non-negligible compared to $\deltaOmegaEcc$, particularly
in highly \sprecing ($\chi_p \sim 1$) and asymmetric ($q \gg 1$)
systems, where it may even dominate over $\deltaOmegaEcc$.  Similar
features can be observed in the dynamical variables of eccentric
\sprecing systems. For example, the \SpEC~\cite{SpECwebsite} code
employs an eccentricity control algorithm based on fitting
$\dotOmegaOrb$, the first derivative of the orbital angular frequency
$\omegaOrb$ computed from the black hole trajectories, where
$\dotOmegaOrb$ is written as a sum of three terms analogous to
equation~(\ref{eq:omega_decomposition}) (see e.g. equation (47) of
~\cite{Buonanno:2010yk}, equation (4)
of~\cite{Habib:2024soh}, and equation (5) of~\cite{Knapp:2024yww}).

In this section, we focus on an example NR simulation \sxsID{4438}
with $q=3$ and $\chi_p = 0.8$. The system has an initial eccentricity
$\egw \approx 5\times 10^{-3}$ at $\tStart \approx -6000M$. Due to the
small eccentricity and large \sprec, spin-induced features in
$\omegaGW$ are non-negligible. We discuss procedures for accurately
estimating $\egw$ in such a system by removing spin-induced features
from the waveform quantities.

\subsubsection{Identifying the timescales}
\label{sec:Identifying_the_timescales}
\begin{figure}
  \centering
  \includegraphics[width=\textwidth]{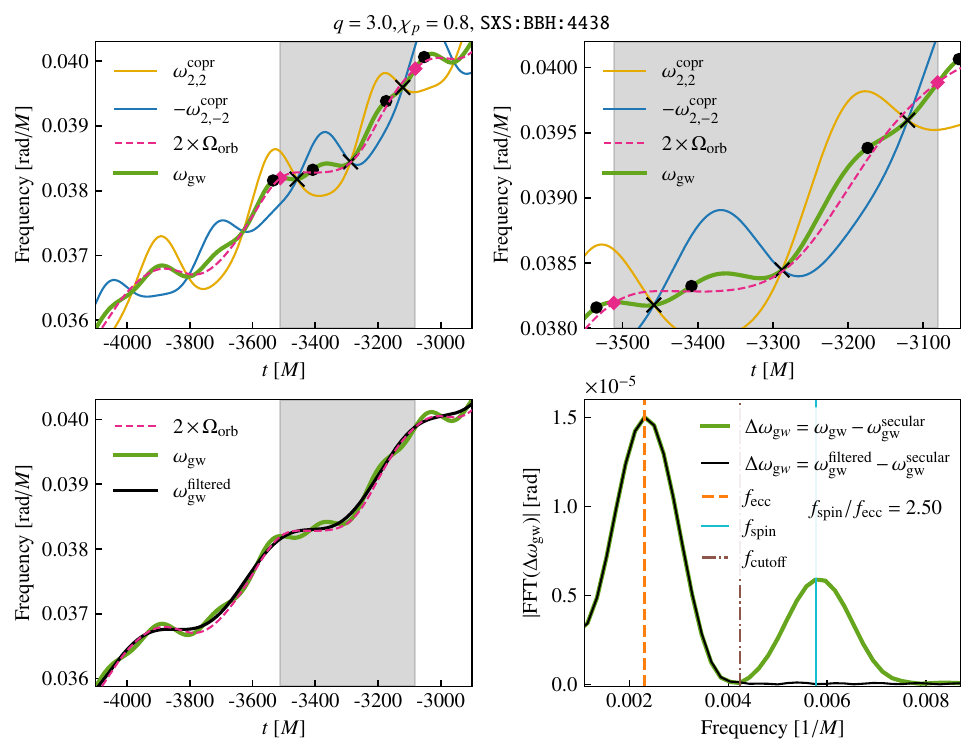}
  \caption{Spin-induced modulations in $\omegaGW$ for a
low-eccentricity, high-\sprec system \sxsID{4438}. {\itshape
Top-left:} $\omegaGW$, defined as the antisymmetric combination of
$\omegaCopr_{2,2}$ and $\omegaCopr_{2,-2}$, exhibits orbital-scale
oscillations due to eccentricity and faster spin-induced
modulations. The eccentricity-induced oscillations resemble those in
the orbital angular frequency $\omegaOrb$. The shaded region spans one
orbit, bounded by two consecutive pericenters (pink diamonds) in
$\omegaOrb$. {\itshape Top-right:} Zoom-in of the shaded region. The
shaded region (one eccentric oscillation) contains $\sim 2.5$
spin-induced oscillations; each oscillation is marked by two
consecutive local peaks (dark circles) in $\omegaGW$.  As expained in
section~\ref{sec:Identifying_the_timescales}, these spin-induced
oscillations be associated with the intersections of
$\omegaCopr_{2,2}$ and $\omegaCopr_{2,-2}$ (crosses).
{\itshape Bottom-right:} Amplitude spectra of
$(\omegaGW - \omegaGWSecular)$ and $(\omegaGWFiltered -
\omegaGWSecular)$ reveal the frequency components. The high-frequency
mode associated with spin-induced oscillations is suppressed after
filtering. {\itshape Bottom-left:} Spin-induced modulations are
removed using a low-pass filter with a cutoff frequency ($\fCutoff$)
located at the trough between the eccentric mode ($\fEcc$) and the
spin-induced mode ($\fSpin$) in the amplitude spectrum shown in the
{\itshape Bottom-right} panel.}
\label{fig:low_ecc_high_prec}
\end{figure}
To remove the spin-induced oscillations from $\omegaGW$, we need to
identify the timescale of these oscillations. We can achieve this by
either looking at the local extrema in these modulations in time
domain or looking at the amplitude spectra of the waveform modes in
the frequency domain.

The {\itshape top-left} panel of figure~\ref{fig:low_ecc_high_prec}
shows $\omegaGW$ of \sxsID{4438}. For comparison, it also shows
$\omegaCopr_{2,2}$, $\omegaCopr_{2,-2}$ and $2\times\omegaOrb$. We use
the time between two consecutive peaks in these frequencies to
estimate the characteristic timescale of the modulations in
them. These ``peaks'' and ``troughs'' are obtained after subtracting
the secular trend from the respective frequency (using a method
similar to the \mFreqFits
method from~\cite{Shaikh:2023ypz}): for this reason, the peaks
and troughs that are identified in figure~\ref{fig:low_ecc_high_prec}
may not correspond exactly with the points one might associate by eye
as turnover points.

The area around the shaded region in the {\itshape top-left} panel of
figure \ref{fig:low_ecc_high_prec} is zoomed-in in the {\itshape
top-right} panel. The shaded region spans one orbit bounded by the two
local peaks (pink diamonds) of $\omegaOrb$. The amplitude of the
oscillation in $\omegaCopr_{2,2}$ and $\omegaCopr_{2,-2}$ are much
larger compared to the same in $2\times\omegaOrb$. $\omegaGW$ shows
oscillations of smaller amplitude compared to those in
$\omegaCopr_{2,2}$, and closely follows the \orbt oscillations due to
eccentricity in $2\times\omegaOrb$. However, some spin-induced
modulations are visible in $\omegaGW$: a local trough forms in
$\omegaGW$ near the intersections (crosses) of
$\omegaCopr_{2,2}$ and $\omegaCopr_{2,-2}$, leading to corresponding
local peaks (dark circles) in $\omegaGW$ between these intersection
points. We associate these shorter timescale oscillations with
spin-induced effects ($\deltaOmegaSpin$ in
equation~(\ref{eq:omega_decomposition}) ), as they arise from
asymmetries between $\omegaCopr_{2,2}$ and $\omegaCopr_{2,-2}$.

Henceforth, we refer to frequency of the spin-induced oscillations in
$\omegaGW$ (i.e. due to $\deltaOmegaSpin$) as $\fSpin$, and the
frequency of the eccentricity-induced oscillations (i.e. due to
$\deltaOmegaEcc$) as $\fEcc$. In figure~\ref{fig:low_ecc_high_prec},
one eccentric oscillation (shaded region between the two pink
diamonds) contains $\sim 2.5$ spin-induced oscillations, where one
spin-induced oscillation spans two consecutive local peaks indicated
by dark circles. Therefore, the frequency $\fSpin$ is approximately
$2.5\times \fEcc$. We can verify the relation between $\fEcc$ and
$\fSpin$ by looking at the the amplitude spectrum of the
residual frequency $\Delta\omegaGW = \omegaGW - \omegaGWSecular$
(obtained using a method similar to \mFreqFits) in the {\itshape
bottom-right} panel of figure \ref{fig:low_ecc_high_prec}. The
amplitude spectrum has two modes at the two frequencies $\fEcc$ and
$\fSpin$ related to the eccentric and the spin-induced oscillations,
respectively, with $\fSpin/\fEcc = 2.5$ consistent with the inference
from the timescales in the {\itshape top-right} panel. In general,
across the inspiral, we find that $\fSpin/\fEcc \sim 2 - 3$ which is
consistent with \cite{Habib:2024soh} (see their figure 3) although in
\cite{Habib:2024soh} the oscillations are in $\dotOmegaOrb$. To
  summarize, $\fSpin$ and $\fEcc$ can be extracted in two ways: either
  from the local peaks of $\Delta\omega(t)$ in the time domain, or
  from the peaks in its amplitude spectrum. In the remainder of this
  work, we refer to the values of $\fSpin$ and $\fEcc$ obtained from
  the amplitude spectrum.

\subsubsection{Removing spin induced effects using low-pass filter}
\label{sec:Remove_spin_induced_effects_using_lowpass_filter}
In the previous section, we noticed that $\fSpin \approx n \fEcc$,
where $n$ varies between $2$ to $3$ across the inspiral. Therefore,
one can remove the spin-induced oscillations by applying a low-pass
filter on $\Delta\omegaGW$ with the cutoff frequency $\fCutoff$ chosen
appropriately such that $\fEcc < \fCutoff < \fSpin$. Such a method has
been applied to remove the spin-induced oscillation from
$\dotOmegaOrb$ in~\cite{Habib:2024soh}. We follow a similar procedure
as outlined in section II.B.1 of~\cite{Habib:2024soh}.

Because the orbital period decreases over time as the binary
inspirals, using a fixed $\fCutoff$ to perform low-pass filter for the
entire $\omegaGW$ may not always be effective, especially for longer
waveforms.  Therefore, we need to perform the low-pass filter on
$\omegaGW$ segment-wise, where each segment may span a few orbits such
that both eccentric and spin-induced oscillations are captured within
the segment, and at the same time, $\fSpin$ and $\fEcc$ remains nearly
constant across the segment.  If $\Delta T(t)$ is the timescale of the
spin-induced modulations at $t$, then a segment of size $10 \times
\Delta T(t)$, centered around $t$, would contain $\sim 3-5$ orbital
cycles since one orbital cycle contains $\sim 2-3$ spin-induced
oscillations (as discussed above).  As discussed in the previous
section, $\Delta T$ can be estimated by finding the time difference
between consecutive local peaks in $\Delta\omegaGW$. However, we
employ an easier alternative approach to find $\Delta T$ by looking at
the time difference between the consecutive intersection points of
$\omegaCopr_{2,2}$ and $\omegaCopr_{2,-2}$ since the intersection
points can be associated with the local troughs in $\omegaGW$ (see
figure~\ref{fig:low_ecc_high_prec}).

Assuming there are $N+1$ such intersection points, we compute the gaps
$\{\Delta T_i = T_{i+1} - T_i\}_{i=1,..,N}$ between the intersection
points $\{T_i\}_{i=1,..,N+1}$, and associate each $\Delta T_i$ with
the time at the midpoint $\{t_i = (T_{i+1} +
T_{i})/2\}_{i=1,..,N}$. We use these data points $\{(t_i, \Delta
T_i)\}_{i=1,..,N}$, to construct an interpolant $\dtIntersect(t)$ for
the period of the occurrences of the intersection points. An
interpolant for the frequency of spin-induced oscillations is then
obtained as $\fSpinIntersect(t) = 1 / \dtIntersect(t)$. Note that we
do not use $\fSpinIntersect(t)$ directly for low-pass
filtering. Instead, it serves as an initial estimate to guide the
accurate extraction of $\fSpin$ from the amplitude spectrum of
$\Delta\omegaGW(t)$ (step \ref{it:fspin_from_amp_spectrum} below).

We use the following steps to filter out the spin-induced oscillations
in $\omegaGW$ and obtain the filtered frequency $\omegaGWFiltered$
containing only eccentricity induced oscillations:

\begin{enumerate}[label=(\alph*)]
\item To ensure that we are using a suitable $\fCutoff$, we process
  the data using overlapping segments.  The overlapping segments are of
  a varying size set as $10\times\dtIntersect(t_s)$, where $t_s$ is the
  time at the center of the segment. The times $\{t_s\}$ are selected at
  regular intervals of size $5\times\min(\dtIntersect(t))$ to ensure
  some overlap between successive segments.
  
\item \label{it:fspin_from_amp_spectrum} For each segment, we compute
  the residual $\Delta \omegaGW(t)$. We compute the Fast Fourier
  transform of $\Delta \omegaGW(t)$ and denote it as
  $\mathrm{FFT}({\Delta\omegaGW})(f)$. We search for the modes related
  to the eccentric and spin-induced oscillations in the amplitude
  spectrum $\left|\mathrm{FFT}({\Delta\omegaGW})\right|(f)$ by locating
  the local frequency peaks using \texttt{scipy.signal.find\_peaks} from
  \texttt{Scipy}~\cite{2020SciPy-NMeth}.  To locate $\fEcc$, we restrict
  the search to the range $(1/4)\fSpinIntersect(t_{\mathrm{mid}}) \leq
  \fEcc \leq (1/2) \fSpinIntersect(t_{\mathrm{mid}})$, where
  $t_{\mathrm{mid}}$ is the midpoint of the current segment and we use
  the $\fSpinIntersect(t)$ interpolant obtained earlier. Similarly, we
  restrict the search for $\fSpin$ 
  to the range $\fSpin \geq (3/4) \fSpinIntersect(t_{\mathrm{mid}})$.
  See {\itshape bottom-right} panel of figure
  \ref{fig:low_ecc_high_prec} for an example, where we can see two
  distinct peaks: the peak at higher frequency corresponds to $\fSpin$
  and the peak at the lower frequency corresponds to $\fEcc$.

\item We then set the cut-off
  frequency $\fCutoff$ to the frequency at the local trough between
  $\fEcc$ and $\fSpin$. We apply a low-pass filter and set
  $\left|\mathrm{FFT}({\Delta\omegaGW})\right|$ to zero for
  frequencies greater than $\fCutoff$ and take inverse Fourier
  transform to obtain the filtered residual $\Delta
  \omegaGWFiltered$. Finally, we get the filtered $\omegaGW$ as
  $\omegaGWFiltered = \Delta\omegaGWFiltered + \omegaGWSecular$, where
  $\omegaGWSecular$ is secular trend in $\omegaGW$.
  The {\itshape bottom-left} panel of figure~\ref{fig:low_ecc_high_prec}
  shows the filtered frequency
  $\omegaGWFiltered$ which closely resembles $2\times\omegaOrb$. The
  {\itshape bottom-right} panel shows the amplitude spectrum of the
  residual of the filtered frequency, that is, $(\omegaGWFiltered -
  \omegaGWSecular)$ which no longer contains the modes at $\fSpin$
  present in the unfiltered case.

\item The filtered overlapping segments are combined using a blending function to
  obtain the full filtered time series. We use \texttt{numpy.hanning} from the
  \texttt{Numpy} library for smooth blending.
\end{enumerate}

So far, we have described the spin-induced oscillations in
  $\omegaGW$. Similar features also appear in the amplitude
  $\ampGW$. Since amplitude based methods, like the \mAmpFits, locate
  the pericenters and apocenters using the local extrema in $\ampGW$ or
  quantities derived from $\ampGW$, it is necessary to remove the
  spin-induced oscillations from $\ampGW$ as well. The procedure to
  remove the spin-induced oscillations from $\ampGW$ is the same as
  described above for $\omegaGW$.

\subsubsection{When to apply filtering}
\label{sec:When_to_perform_filtering}
\begin{figure}[h]
  \centering
  \includegraphics[width=\textwidth]{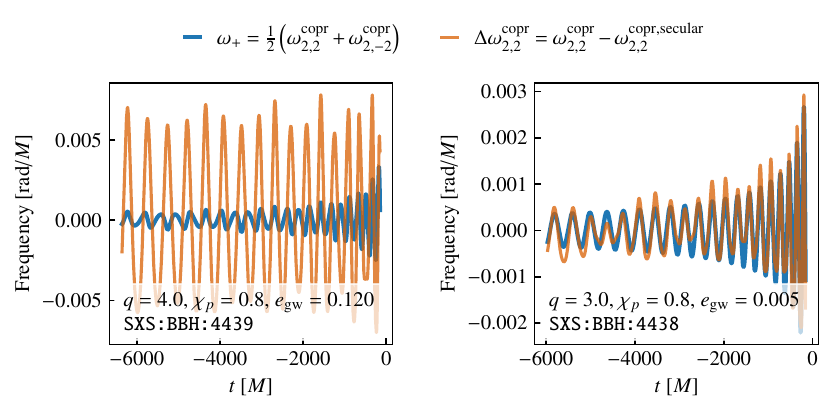}
  \caption{{\bfseries Requirement for filtering.} {\itshape Left:}
    $\omegaAsymmetry$ is smaller compared to the residual coprecessing
frequency $\Delta\omegaCopr_{2,2}$, and, therefore, filtering
$\omegaGW$ is not required. {\itshape Right:} Amplitude of the
$\omegaAsymmetry$ is nearly the same as the amplitude of
$\Delta\omegaCopr_{2,2}$. In this case, filtering is required to remove the
non-negligible spin-induced effect.}
  \label{fig:requirement_for_filtering}
\end{figure}
Filtering is required to remove spin-induced oscillations and obtain
oscillations caused solely by eccentricity. However, the procedure
explained in the previous section should be applied only in the small
eccentricity regime. This is because at high eccentricity, it can be
challenging to distinguish between modes due to eccentric higher
harmonics~\cite{Yunes:2009yz,Islam:2025rjl,Islam:2025llx}
(which appear as integer multiples of $\fEcc$) and the mode due
to spin-induced oscillation (which appear as $\fSpin \sim 2-3
  \times \fEcc$ as we saw in figure~\ref{fig:low_ecc_high_prec}). Removing higher eccentric harmonics by incorrectly 
identifying as spin-induced oscillation may cause significant bias in 
the measured eccentricity. Also, we do not require filtering at high
eccentricity anyway, because the eccentric modulations become
significantly larger than the spin-induced modulations.

The frequency $\omegaGW$ in equation~(\ref{eq:omega_gw}) is the
antisymmetric combination of $\omegaCopr_{2,2}$ and $\omegaCopr_{2,-2}$
that nearly cancels out the \masym. On the other hand, the symmetric
combination of $\omegaCopr_{2,2}$ and $\omegaCopr_{2,2}$~\cite{Boyle:2014ioa,Blackman:2017pcm,Blackman:2017dfb,Varma:2019csw},
\begin{equation}
  \label{eq:omega_asym}
  \omegaAsymmetry = \frac{1}{2}\left(\omegaCopr_{2,2} + \omegaCopr_{2,-2}\right),
\end{equation}
captures the \masym. Using equation~(\ref{eq:omega_asym}), $\omegaGW$
in equation~(\ref{eq:omega_gw}) can be rewritten as 
\begin{eqnarray}
  \label{eq:omega_gw_with_asym}
  \omegaGW= \omegaCopr_{2,2} - \omegaAsymmetry
\end{eqnarray}
Removing the secular trend from $\omegaGW$, the above equation becomes
\begin{eqnarray}
  \label{eq:residual_with_asym}
  \Delta\omegaGW & = (\omegaCopr_{2,2} - \omegaGWSecular) -
  \omegaAsymmetry.\\
  & \approx (\omegaCopr_{2,2} - \omega_{2,2}^{\mathrm{copr,secular}}) -
    \omegaAsymmetry,
  \label{eq:residual_with_asym_approx}
\end{eqnarray}
where in equation~(\ref{eq:residual_with_asym_approx})
we have assumed that $\omegaGWSecular \approx
\omega_{2,2}^{\mathrm{copr,secular}}$. This is sufficient as we only
need the scale of the two terms, $(\omegaCopr_{2,2} -
\omega_{2,2}^{\mathrm{copr,secular}})$ and $\omegaAsymmetry$, to
estimate how large spin-induced effects are compared to eccentric
effects. Using $\Delta\omegaCopr_{2,2} = \omegaCopr_{2,2} -
\omega_{2,2}^{\mathrm{copr,secular}}$, $\Delta\omegaGW$ can be
approximately written as the sum of two contributions:
\begin{equation}
  \label{eq:residual_in_copr_and_asym}
  \Delta\omegaGW \approx \Delta\omegaCopr_{2,2} - \omegaAsymmetry.
\end{equation}
In equation~(\ref{eq:residual_in_copr_and_asym}), the first
term on the right contains the modulations due to eccentricity as well
as \sprec, whereas the second term is caused solely by \sprec. For
small eccentricity, $\Delta\omegaCopr_{2,2}$ is dominated by the
spin-induced effects, and becomes comparable in magnitude to
$\omegaAsymmetry$. Therefore, we use a threshold on the ratio
$\rFilter$, defined as the ratio of the maximum values of
$\omegaAsymmetry$ and $\Delta\omegaCopr_{2,2}$, to decide whether
filtering is necessary. Since both quantities exhibit oscillations
over time, we compute the average of their values at their respective
local maxima (denoted by superscript ``m'' below),
\begin{equation}
  \label{eq:rfilter}
  \rFilter =
  \frac{\overline{\omegaAsymmetry}}{\overline{\Delta\omegaCopr_{2,2}}},\qquad
  \overline{X} = \frac{1}{N} \sum_{i=1,..,N} X_i^{\mathrm{m}}.
\end{equation}
Both $\Delta\omegaCopr_{2,2}$ and
$\omegaAsymmetry$ oscillate in time about zero, and, hence have
positive values at the local maxima, which ensures that $\rFilter > 0$.
We recommend applying the filter only if $\rFilter > 0.2$. For smaller
$\rFilter$, the spin-induced oscillations are negligible compared to
the eccentricity-induced oscillations, and filtering may accidentally
remove higher eccentric harmonics instead, leading to bias in the
measured eccentricity.
In figure~\ref{fig:requirement_for_filtering}, we compare
$\omegaAsymmetry$ and $\Delta\omegaCopr_{2,2}$ for two example
cases. The {\itshape left} panel shows that the amplitude of
$\omegaAsymmetry$ is small compared to that of
$\Delta\omegaCopr_{2,2}$ for a system with $q=3, \chi_{p}=0.8,
\egw=0.12$. For this system, $\rFilter \approx 0.15$, and filtering
$\omegaGW$ is not recommended. The {\itshape right} panel shows that
the amplitude of $\omegaAsymmetry$ is nearly the same as the
amplitude of $\Delta\omegaCopr_{2,2}$ for a system with $q=3,
\chi_{p}=0.8, \egw=0.005$. For this system, $\rFilter \approx 0.82$
and filtering $\omegaGW$ is recommended.

\subsection{Summary}\label{sec:summary}
\begin{figure}
  \centering
  \includegraphics[width=\textwidth]{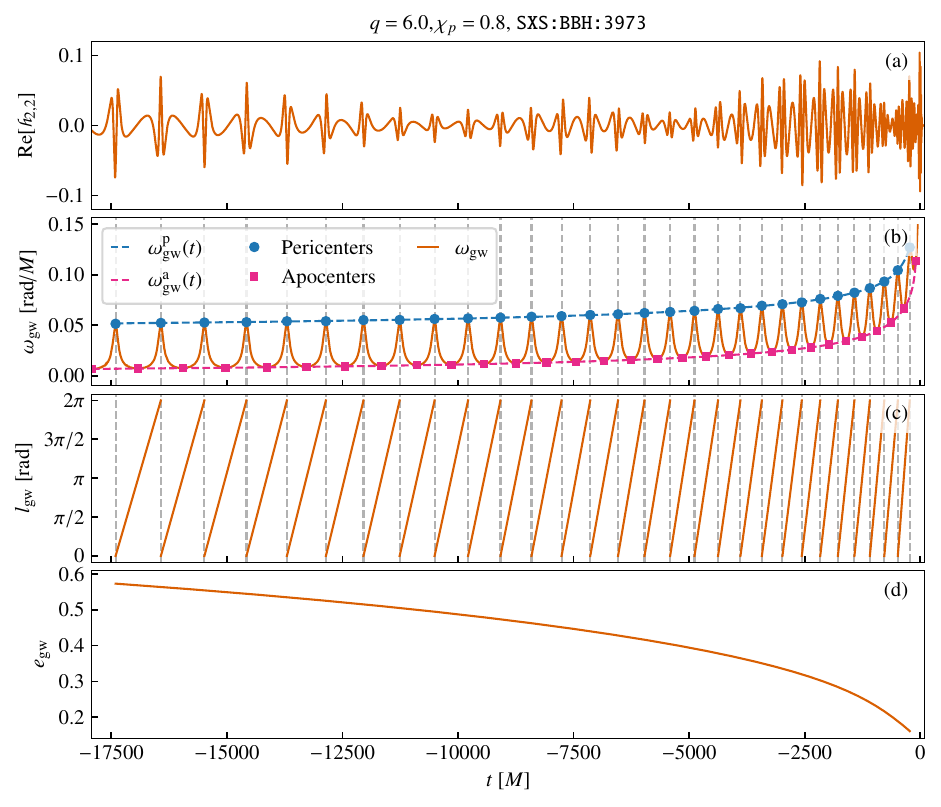}
  \caption{{\bfseries Demonstration of eccentricity $\egw$ and mean
      anomaly $\lgw$ measurement for a strongly spin-precessing system.}
    (a) Real part of the \dm mode in the inertial frame, showing the
    effects of eccentricity and \sprec.
    (b) It shows the frequency $\omegaGW$, antisymmetric combination
    of the $\omegaCopr_{2,2}$ and $\omegaCopr_{2,-2}$. $\omegaGWP(t)$ and
    $\omegaGWA(t)$ are the interpolants through $\omegaGW$ evaluated at
    the pericenters (blue circles) and the apocenters (pink squares),
    respectively. The vertical gray dashed lines denote the pericenter
    times. (c) Time evolution of mean anomaly $\lgw$. $\lgw$ grows
    linearly from 0 to $2\pi$ between successive pericenters
    (equation~(\ref{eq:mean_anomaly})). (d) Time evolution of the
    eccentricity $\egw$ computed using equation~(\ref{eq:egw}) given
    $\omegaGWP(t)$ and $\omegaGWA(t)$.
    }\label{fig:egw_demo}
\end{figure}
\begin{figure}
\centering

\includegraphics[width=\textwidth]{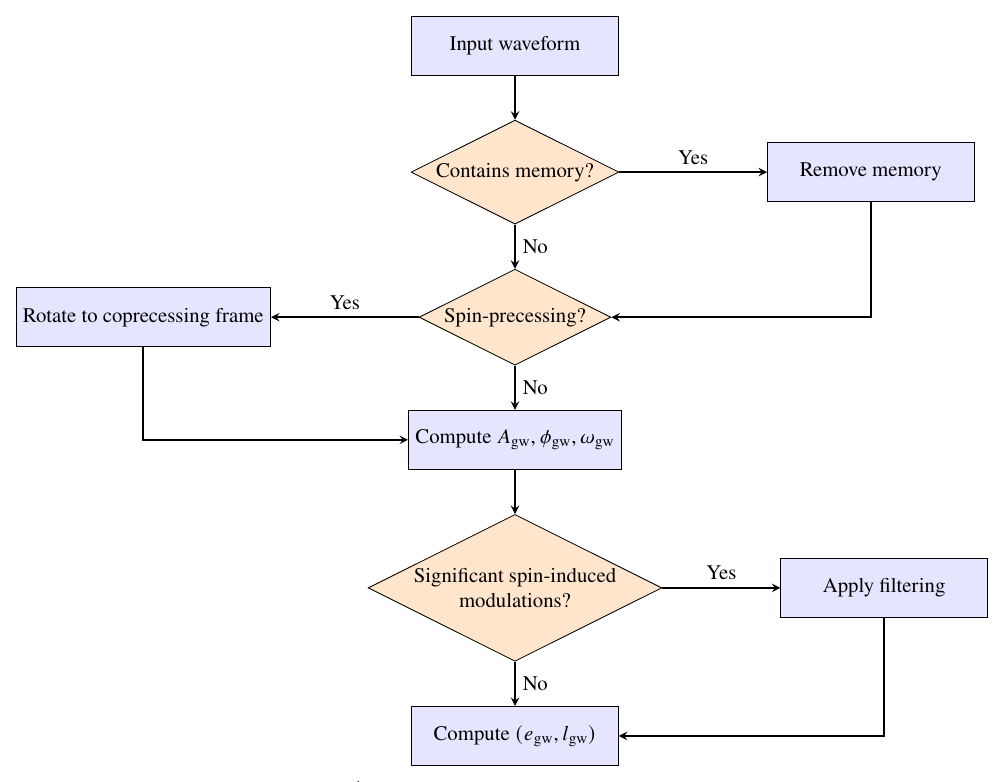}
\caption{{Flowchart illustrating the pipeline for computing the
eccentricity and mean anomaly for a given {eccentric \sprecing} waveform. These
steps are in Section~\ref{sec:summary}.}}\label{fig:flowchart}
\end{figure}

In this section, we summarize the steps to measure eccentricity $\egw$ and mean
anomaly $\lgw$ given a waveform.
{
These steps are also illustrated in the flowchart in
figure~\ref{fig:flowchart}.
}

\begin{enumerate}[label=(\alph*)]
\item If the waveform includes memory contribution, the memory
  contribution is removed from the waveform before it is used for
  eccentricity measurement.
\item \label{it:fist_step} If the system is spin-precessing, the
  waveform is transformed to the coprecessing frame. For \aspin
  systems, the inertial and coprecessing frames are equivalent.
\item Instead of using only the inertial frame \dm mode variables, variables with
  reduced spin-induced oscillation, that is, $\ampGW, \phaseGW$ and
  $\omegaGW$ provided by equation~(\ref{eq:amp_gw}),
  (\ref{eq:phase_gw}) and (\ref{eq:omega_gw}), respectively, are
  used.
\item For small eccentricity and large \sprec, $\ampGW$ and $\omegaGW$
  are  low-pass filtered to remove spin-induced oscillations. Whether 
  filtering is required is decided using the criterion discussed in
  section~\ref{sec:When_to_perform_filtering}.
\item Finally, $\egw$ and $\lgw$ defined in equation~(\ref{eq:egw}) and
(\ref{eq:mean_anomaly}), respectively, are computed using the GW
variables computed in the previous steps. For building $\omegaGWP(t)$
and $\omegaGWA(t)$, using \mRatFit (see section~\ref{sec:rational_fits_for_omega_extrema}) is recommend over \mSpline as
\mRatFit is more robust in providing consistent monotonically
decreasing $\egw(t)$.
\end{enumerate}

Figure~\ref{fig:egw_demo} demonstrates the computation of the
eccentricity $\egw$ and the mean anomaly $\lgw$ using the waveform
from an NR simulation with significant \sprec ($\chi_p = 0.8$) and
large eccentricity (initial $\egw\sim0.57$). Here, we have removed the
memory from the waveform, transformed the waveform to coprecessing
frame, and used \mRatFit for building $\omegaGWP(t)$ and
$\omegaGWA(t)$ (for this case low-pass filtering was not recommended
by the criterion in section~\ref{sec:When_to_perform_filtering}).

Next, we demonstrate the computation of $\egw$ for a system with
significant \sprec ($\chi_p = 0.8$) and small eccentricity (initial
$\egw \sim 0.005$). For this case, low-pass filtering is recommended
by the criterion in section~\ref{sec:When_to_perform_filtering}
to remove spin-induced oscillations from $\omegaGW$
and $\ampGW$. In
figure~\ref{fig:egw_steps_low_ecc_high_prec}, we show how the 
measured $\egw(t)$ varies depending on the methods used for the
measurement. It contains the following cases:
\begin{figure}
  \centering
  \includegraphics[width=\textwidth]{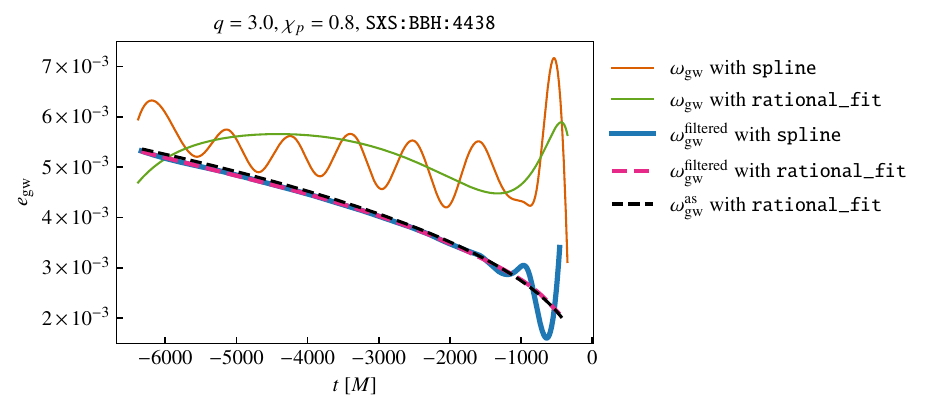}
  \caption{$\egw$ for small eccentricity and high \sprec system
\sxsID{4438}. $\egw$ obtained using $\omegaGW$ without smoothing
filter shows incorrect behavior (oscillates and increases near merger)
which is fixed when using $\omegaGWFiltered$ with \mRatFit for $\egw$
measurement. $\egw$ using $\omegaGWFiltered$ closely matches the
$\egwAS$ of the \aspin counterpart obtained using \SEOB. For
details, see the main text.}
  \label{fig:egw_steps_low_ecc_high_prec}
\end{figure}

\begin{enumerate}[label=(\Alph*)]
\item $\omegaGW$ with \mSpline: \label{enum:mSpline}
This line represents $\egw(t)$ computed using \mSpline interpolants
with unfiltered $\omegaGW$. The spin-induced modulations affect the
locations of local extrema and the values of $\omegaGWP$ and
$\omegaGWA$, which is reflected in the oscillatory behavior of
$\egw$. Additionally, note the increasing trend near the merger, which
is due to the enhanced spin-induced modulation in this region. Unlike
eccentric modulations, spin-induced modulations increase over time.

\item $\omegaGW$ with \mRatFit: \label{enum:mRatFit} Same
as~\ref{enum:mSpline}, but using \mRatFit interpolants. The small
oscillations present in $\egw$ based on \mSpline are gone, but the
overall trend remains incorrect.

\item $\omegaGWFiltered$ with \mSpline: Same
as~\ref{enum:mSpline}, but using $\omegaGWFiltered$. We observe that
$\egw$ is monotonically decreasing for most part of its evolution
except near the merger. Near the merger oscillatory behavior appears
due to the use of \mSpline interpolants as discussed in
section~\ref{sec:rational_fits_for_omega_extrema} and ~\ref{sec:rational_fit}.

\item $\omegaGWFiltered$ with \mRatFit \label{enum:mRatFitPlusFliter}:
Same as~\ref{enum:mRatFit}, but using $\omegaGWFiltered$. Now the
$\egw(t)$ has expected monotonically decreasing behavior over time and
does not have oscillatory behavior near the merger unlike the case for
\mSpline.

\item $\omegaGWAS$ with \mRatFit: To assess the evolution of $\egw(t)$
obtained in~\ref{enum:mRatFitPlusFliter} using $\omegaGWFiltered$, we
also plot $\egwAS(t)$ from the \aspin counterpart using
\SEOB~\cite{Gamboa:2024hli,Gamboa:2024imd} waveform. The \aspin
waveform is generated using the same parameters as the \sprecing
system, except for the spins, which we replace by only the
$z$-components in the case of the \aspin waveform (see
section~\ref{sec:comparison_to_aligned_spin_counterpart} for more
details). Because the waveforms in the coprecessing frame look similar
to the corresponding \aspin waveforms, we expect that the $\egw(t)$
will also look similar to the $\egwAS(t)$ of the \aspin
counterpart. We find that $\egwAS(t)$ closely matches $\egw(t)$, as
expected, but it is necessary to use $\omegaGWFiltered$ to obtain the
correct $\egw(t)$.
\end{enumerate}

\section{Robustness checks}\label{sec:results}
In this section, we check the robustness of our method of measuring
eccentricity discussed in the previous section. In
section~\ref{sec:applications_to_nr_and_eob_waveforms}, we showcase
the applicability of our method on NR waveforms of different
eccentricities and \sprec. In section~\ref{sec:smoothness_tests}, we
put our method to a couple of smoothness tests using a set of \TEOB
waveforms. In
section~\ref{sec:comparison_to_aligned_spin_counterpart}, we compare
the $\egw$ evolution of a set of \sprecing waveforms to their \aspin
counterpart. Finally, in
section~\ref{sec:applicability_in_the_small_eccentricity_and_large_sprec_regime},
we check the robustness of our method when applying to waveforms with
small eccentricity and large \sprec.

\subsection{Applications to NR waveforms}\label{sec:applications_to_nr_and_eob_waveforms}
\begin{figure}
  \centering
  \includegraphics{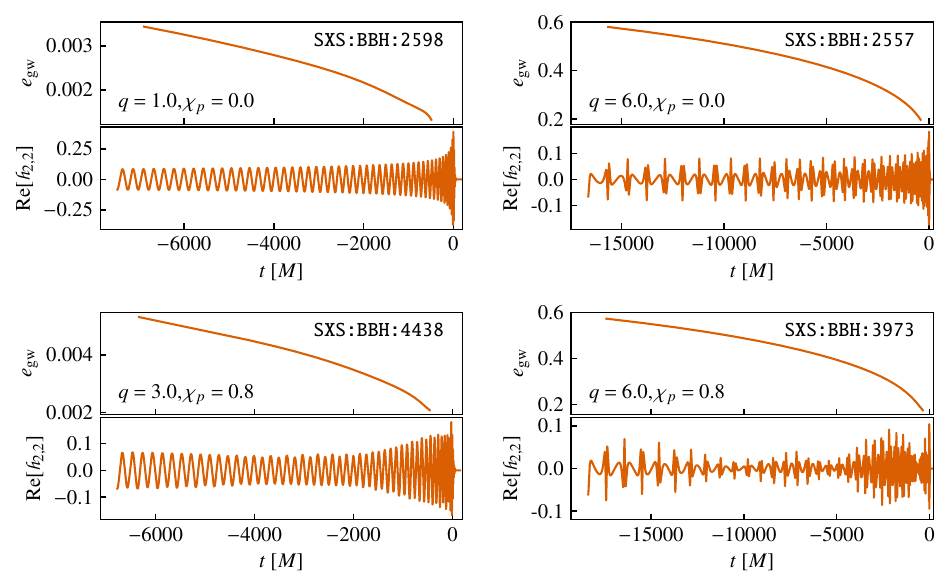}
  \caption{Application to four representative \SXS waveforms obtained
using \SpEC. The {\itshape top} row presents
\aspin eccentric systems, with small eccentricity on the
{\itshape left} and large eccentricity on the {\itshape right}. The
{\itshape bottom} row presents similar plots for \sprecing eccentric
systems—small eccentricity with high \sprec on the {\itshape left} and
high eccentricity with high \sprec on the {\itshape right}. For each
case, the {\itshape top} panel shows the $\egw$ evolution, and the
{\itshape bottom} panel shows the real part of the corresponding $\h_{2,2}$.
}
  \label{fig:application_on_sxs}
\end{figure}

\begin{figure}
  \centering
  \includegraphics{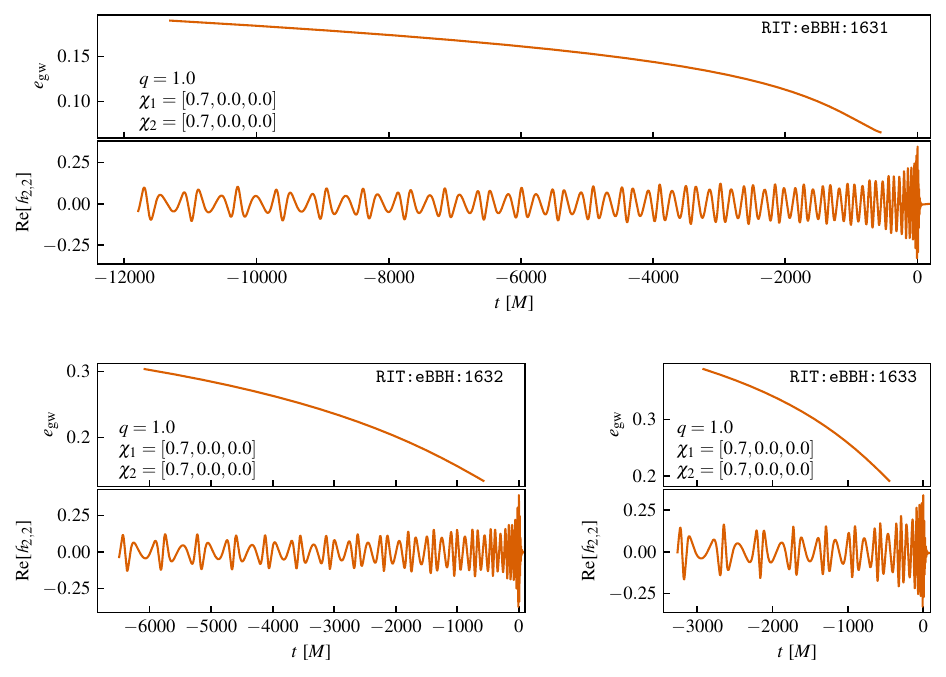}
  \caption{Application to \texttt{RIT} eccentric \sprecing
waveforms. For each case, the {\itshape top} panel shows the $\egw$
evolution and the {\itshape bottom} panel shows the real part of the
corresponding $\h_{2,2}$.
}
  \label{fig:application_on_RIT}
\end{figure}

In this section, we apply our method to compute $\egw$ from a set of NR
waveforms. First, we apply it on waveforms from binary black hole systems
simulated using \SpEC~\cite{SpECwebsite}.  We select four representative cases,
and plot the resulting $\egw$ in figure~\ref{fig:application_on_sxs}: low
eccentricity with \aspins ({\itshape top-left}), high eccentricity with \aspins
({\itshape top-right}), low eccentricity with high \sprec ({\itshape
bottom-left}), and high eccentricity with high \sprec ({\itshape
bottom-right}). Next, in figure~\ref{fig:application_on_RIT}, we repeat this
for a few NR waveforms from the
\texttt{RIT}~\cite{RITCatalog,Healy:2020vre,Healy:2022wdn} catalog. We select
three simulations for this purpose, which contain a sufficient number of orbits
for robust $\egw$ estimation~\footnote{{{As noted {previously} in~\cite{Shaikh:2023ypz},
we find that $\egw(t)$ near the merger may become nonmonotonic as it becomes
harder to define an orbit in this regime. To avoid this nonmonotonic behavior,
by default, we discard the last two orbits of the waveform before computing
$\egw(t)$. We require at least two orbits in the remaining part of the waveform
to successfully build an interpolant. Therefore,} this method to compute
eccentricity from the waveform requires at least $\roughly 4-5$ number of
orbits~\cite{Shaikh:2023ypz}.}}. The RIT simulations share the same mass ratio ($q=1$) and spin
parameters ($\vchi_1=[0.7,0,0], \vchi_2=[0.7,0,0]$).  In each subplot of
figure~\ref{fig:application_on_sxs} and figure~\ref{fig:application_on_RIT},
the {\itshape top} panel shows the $\egw$ evolution, while the {\itshape
bottom} panel presents the real part of the corresponding $\h_{2,2}$. These
cases illustrate the applicability of our method to systems with varying
eccentricity and \sprec.

\subsection{Smoothness tests}\label{sec:smoothness_tests}
At present, \TEOB~\cite{Gamba:2024cvy,Albanesi:2025txj} is the only
publicly available time-domain waveform model that supports both
eccentricity and \sprec, but it does not include \masym terms. In
other words, $\ampGW, \phaseGW$ and $\omegaGW$ are identical to the
corresponding quantities from the (2, 2) mode in the coprecessing
frame, that is, $\ampCopr_{2, 2}, \phaseCopr_{2,2}$ and
$\omegaCopr_{2, 2}$, respectively. However, unlike the NR waveforms,
\TEOB waveforms are not restricted to discrete points in the parameter
space, and therefore, serve as a useful check of the robustness of our
implementation on a larger set of waveforms. We demostrate that our
method can be applied to \TEOB waveforms robustly by performing
smoothness tests. {As noted in~\cite{Shaikh:2023ypz}, such smoothness
tests are important not just as a robustness check of the $\egw$
measurement method, but also of the underlying waveform model.}
\begin{figure}
  \centering
  \includegraphics[width=\textwidth]{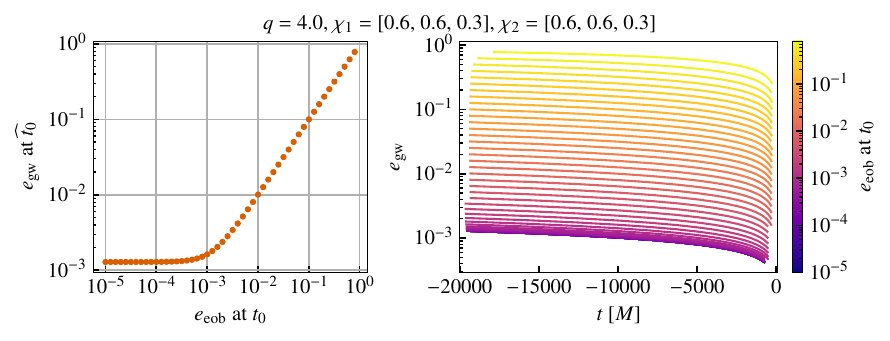}
  \caption{{\bfseries Smoothness tests.} {\itshape Left:} $\egw$ vs
$\eEOB$ smoothness test. The $y$-axis represents the eccentricity
$\egw$ at the earliest time $\tStartHat$ and the $x$-axis represents
the input model eccentricity $\eEOB$ at $\tStart=-20000M$. The
measured eccentricity and the input model eccentricity follows the
$\egw=\eEOB$ line for $\eEOB \gtrsim 10^{-3}$. For $\eEOB \lesssim
10^{-3}$, $\egw$ becomes constant implying that the physical content
of the waveform stops changing even though the model input
eccentricity $\eEOB$ keeps changing.  {\itshape Right:} $\egw$ vs $t$
smoothness test. The colors represent the initial eccentricity $\eEOB$
at $\tStart$ used as an input to \TEOB. $\eEOB$ varies from $10^{-5}$
to 0.8. We choose the starting frequency such that the waveforms start
at $\tStart\approx -20000M$.}
  \label{fig:egw_vs_t}
\end{figure}

To ensure the robustness of our implementation, we put it to a couple
of tests. The first test checks the relation between the measured
$\egw$ and the model input eccentricity $\eEOB$, and the second test
checks the smoothness of the measured $\egw$ evolution. For both
tests, we generate a set of 50 eccentric \sprecing waveforms using the
public waveform model \TEOB~\cite{Gamba:2024cvy}. For generating this
set of waveforms, we vary the initial input $\eEOB$ from $10^{-5}$ to
0.8 while keeping the mass ratio, initial input mean anomaly ($\lEOB$)
and spins fixed at $q = 4.0, \lEOB=\pi, \chi_1=\chi_2=[0.6, 0.6,
0.3]$. We choose the initial input frequency such that the waveforms
start at $\tStart \approx -20000M$.

In figure~\ref{fig:egw_vs_t}, the {\itshape left} panel shows the
relation between the input $\eEOB$ at $\tStart$ and the measured
$\egw$ at $\tStartHat$, the earliest time where $\egw$
can be measured. $\tStartHat$ is determined by $\tStartHat = \max(\tP_0,
\tA_0)$, where $\tP_{0}$ is the time at the first pericenter and
$\tA_{0}$ is the time at the first apocenter.  For
$\eEOB \gtrsim 10^{-3}$, $\egw$ vs $\eEOB$ follows the $\egw=\eEOB$
line. For $\eEOB \lesssim 10^{-3}$, $\egw$ plateaus at $\approx
10^{-3}$. This behavior {was also noted in ~\cite{Shaikh:2023ypz}, and
} implies that the \TEOB model ceases to produce any difference in the
physical content of the waveforms below $\eEOB \lesssim 10^{-3}$. The
{\itshape right} panel shows the evolution of $\egw$. The colors
represent the model input eccentricity $\eEOB$ at $\tStart$.  It shows
the expected monotonically decreasing trend of $\egw$ with time. All
of the $\egw(t)$ curves decay smoothly over time and across change of
initial eccentricity, as expected. We also note that for $\eEOB
\lesssim 10^{-3}$, the $\egw(t)$ curves overlap with each other, for
the same reason noted above.

\begin{figure}
  \centering
  \includegraphics{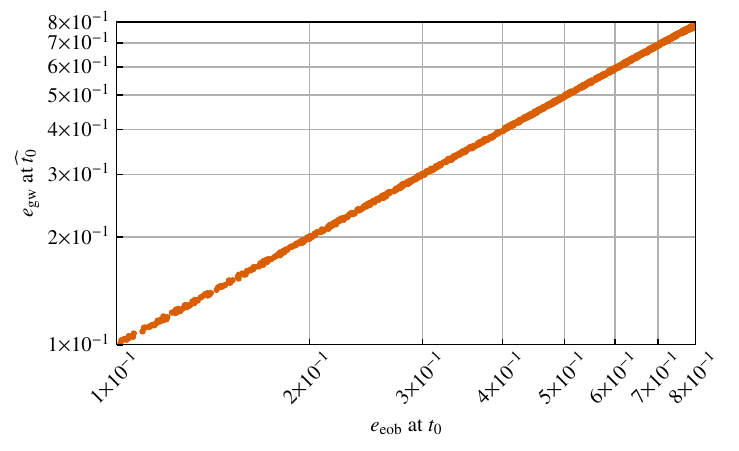}
  \caption{{\bfseries $\egw$ at $\tStartHat$ vs. $\eEOB$ at $\tStart$ across
      the parameter space.} Same as the {\itshape left} panel of figure
    ~\ref{fig:egw_vs_t}, except we now consider a set of 1000 random input parameters
    with varying $(q, \boldsymbol{\chi}_1,\boldsymbol{\chi}_2, \eEOB, \lEOB)$.}
  \label{fig:robustness_across_param_space}
\end{figure}

{
In above smoothness tests, all the parameters are kept fixed except the initial
input eccentricity $\eEOB$. To test the robustness of our method across the
parameter space, we now investigate the relation between the $\eEOB$ at
$\tStart$ and the measured eccentricity $\egw$ at $\tStartHat$ at 1000 set of
input parameters {with varying} $(q, \boldsymbol{\chi}_1,\boldsymbol{\chi}_2,
\eEOB, \lEOB)$. For generating these points in the parameter space, we {sample
randomly from} $q \in \mathcal{U}(1, 10)$, $\eEOB \in \mathcal{U}(0.1, 0.8),
\lEOB \in \mathcal{U}(0, 2\pi)$, $\chi_{1,2}\in \mathcal{U}(0, 1)$,
$\cos\theta_{1,2}\in \mathcal{U}(-1, 1)$ and $\varphi_{1,2}\in \mathcal{U}(0,
2\pi)$, where $\theta_{1,2}$ and $\varphi_{1,2}$ are the polar and the azimuthal
angles, respectively, of $\boldsymbol{\chi}_{1,2}$ at the starting time
$\tStart$.\footnote{$\mathcal{U}(a,b)$ denotes a uniform distribution in the
range $(a, b)$.} We choose the starting frequency such that the waveforms start
at $\tStart \approx -20000M$. Figure~\ref{fig:robustness_across_param_space}
shows $\egw$ at $\tStartHat$ vs. $\eEOB$ at $\tStart$
for these 1000 points. We find good agreement, with the 
percentage difference having a
median value of $0.78\%$ and a 95th percentile value of
1.92\%.
}

\subsection{Comparison to \aspin counterpart}\label{sec:comparison_to_aligned_spin_counterpart}

\begin{figure}[h]
  \centering
  \includegraphics[width=\textwidth]{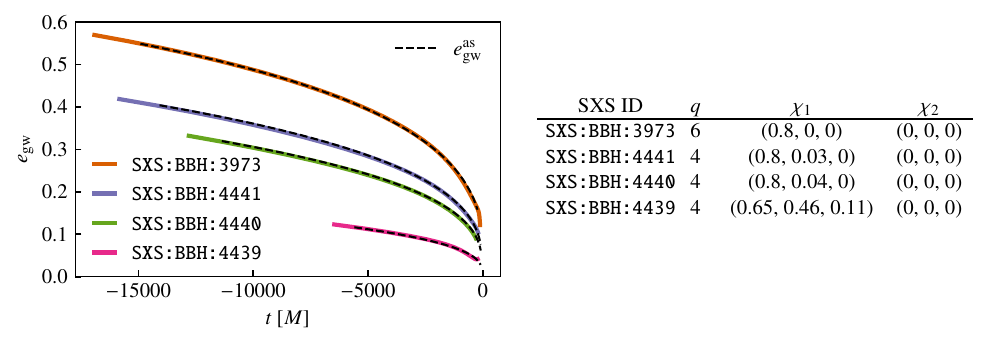}
  \caption{$\egw$ vs $\egw^{\mathrm{as}}$ for eccentric \sprecing NR
waveforms from the \SXS Collaboration. The solid lines represent the
$\egw$ evolution for different NR simulations with varying initial
eccentricity. For each of these lines, the corresponding
$\egw^{\mathrm{as}}$ is shown using dashed line. $\egw^{\mathrm{as}}$
is obtained using \SEOB model with the same mass ratio and
$z$-components of the spins. The eccentricity parameters ($\egw,
\lgw$), at a reference frequency $\freqref$ corresponding to the
second pericenter, are estimated first from the NR waveform, and then
used as an input to \SEOB.}\label{fig:egw_vs_egw_as}
\end{figure}
\begin{figure}[h]
  \centering
  \includegraphics{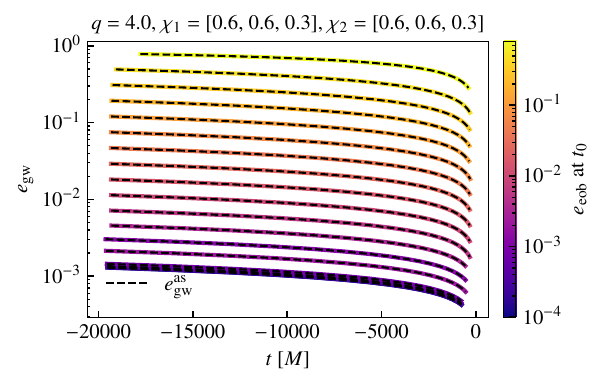}
  \caption{$\egw$ vs $\egw^{\mathrm{as}}$ for \TEOB. Same as
    figure~\ref{fig:egw_vs_egw_as}, but here we use \TEOB to show the
    comparison for a larger set of waveforms with initial $\eEOB$ in
    $[10^{-4}, 0.8]$. In this case, the input parameters for the \sprecing
    and the \aspin waveforms are the same, except that for the
    \aspin case, only $z$ components of the spins are
    non-zero. The colorbar shows the initial input $\eEOB$ at the
    starting time $\tStart \approx -20000M$.}
  \label{fig:dali_egw_vs_egw_as}
\end{figure}

For computing $\egw(t)$ of a \sprecing system, the waveform is
transformed to the coprecessing frame. This transformation removes
most of the \sprec effects from the waveform, making them similar to
those of an \aspin system. Therefore, we expect the measured $\egw(t)$
to also look similar to the eccentricity $\egwAS$ of its \aspin
counterpart. The waveform of the \aspin counterpart is generated using
the same parameters as the \sprecing system, except that the spins are
replaced by only the $z$-components $(\chi_{1z}, \chi_{2z})$. We find
that $\egw$ and the eccentricity of its \aspin counterpart,
$\egw^{\mathrm{as}}$, are very close to each other as expected.

First we perform this on a set of eccentric \sprecing NR simulations
(see ~\cite{NRpaper} for further details on these simulations)
performed using \SpEC~\cite{SpECwebsite}. Because the NR simulations
are performed at discrete points in the binary's parameter space, it
is difficult to find the \aspin counterpart from the existing NR
catalog. Instead, in this case, we generate the \aspin counterpart
using the \SEOB model~\cite{Gamboa:2024imd,Gamboa:2024hli}. As an
input for the model, we use the same mass ratio and $z$-components of
the spins ($\chi_{1z}, \chi_{2z}$) as the NR \sprecing waveform. For
the eccentricity parameters, we first compute the eccentricity $\egw$
and mean anomaly $\lgw$ at the earliest dimensionless frequency
$\freqref${\footnote{$\freqref$ is related to reference time
$t_{\mathrm{ref}}$, where ($\egw, \lgw$) is computed, by the equation
$\langle \omegaGW\rangle(t_{\mathrm{ref}}) = 2\pi \freqref$, where
$\langle \omegaGW\rangle$ is the orbit averaged $\omegaGW$. See
section II.E in~\cite{Shaikh:2023ypz} for more details, where it is
discussed for aligned-spin systems, but applies to \sprecing systems
as well by working with $\omegaGW$, instead of $\omegatwotwo$.}},
after removing 2 orbits\footnote{One orbit is approximated as a
phase difference of $2\pi$ in the orbital phase, which is computed
from the black hole trajectories provided along with the simulation
data.} of data from the initial part of the waveform as junk
radiation. Then we use these as the input {initial eccentricity and
mean anomaly} to the \SEOB model for generating the \aspin
waveform. In figure~\ref{fig:egw_vs_egw_as}, we plot $\egw(t)$ for the
four NR waveforms with varying initial eccentricity. The corresponding
$\egwAS(t)$ of the \aspin counterpart is shown using the dashed lines.

Next, we do the same study using the \TEOB model. In this case, we can
generate the \aspin counterpart using the exact set of input
parameters (mass ratio, eccentricity, mean anomaly and the
$z$-component of the spin vectors) as the \sprecing one. Also, we can
generate a larger set of waveforms with \TEOB. In
figure~\ref{fig:dali_egw_vs_egw_as}, we plot the $\egw(t)$ for 20
\TEOB waveforms with $q=4, \chi_1=\chi_2=[0.6, 0.6, 0.3]$ and initial
$\eEOB$ varying from $10^{-4}$ to $\eEOB=0.8$. We choose the starting
frequencies so that the resulting waveforms start at $\tStart \approx
-20000M$. The input initial $\eEOB$ are shown using the colorbar. The
dashed lines represent the eccentricity $\egwAS(t)$ of the \aspin
counterpart for each of these waveforms.

These checks serve two purposes: First, they demonstrate the
robustness of our method in measuring $\egw$ from \sprecing waveforms,
as $\egw^{\mathrm{as}}$ measured from the \aspin counterpart serves as
a useful reference. Second, even though the association of an \aspin
counterpart to a \sprecing system is only approximate, these checks
demonstrate that the approach of frame-twisting an \aspin waveform to
approximate a \sprecing waveform can be a reasonable waveform modeling
strategy even in the eccentric regime (see \cite{Gamba:2024cvy} for
work in this direction).

\subsection{Applicability in the small eccentricity \& large \sprec
  regime}\label{sec:applicability_in_the_small_eccentricity_and_large_sprec_regime}
\begin{figure}
  \centering
  \includegraphics{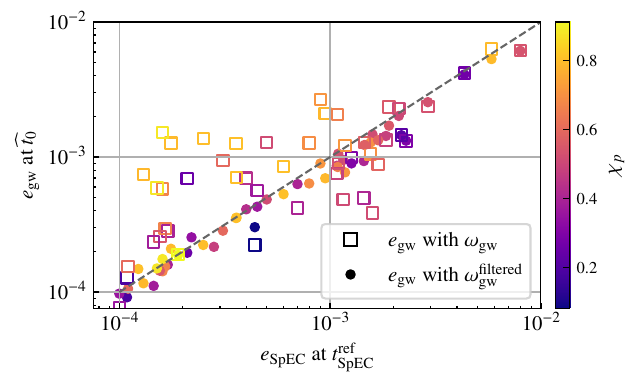}
  \caption{$\egw$ at $\tStartHat$ vs $\eSpEC$ at $\tRefSpEC$ for
\sprecing systems with $10^{-4} \lesssim \egw \lesssim 10^{-2}$. Due
to spin-induced effects dominating over eccentric effects, $\egw$
measured using unfiltered $\omegaGW$ can differ from the corresponding
$\eSpEC$ values by a factor of $\sim 10$ for $\eSpEC \lesssim
10^{-3}$. For the same cases, when using $\omegaGWFiltered$, $\egw$
and $\eSpEC$ agrees within a factor of $\sim 2$. The colors represent
the initial values of $\chi_p$.  One can see that, in general, with
increasing $\chi_p$, filtering becomes more important in the small
eccentricity regime.}
\label{fig:small_nr_eccentricity}
\end{figure}
In section~\ref{sec:low_eccentricity_and_high_precession}, we
discussed how spin-induced effects may become non-negligible compared
to the eccentric modulations in the small eccentricity and large
\sprec limit, and provided a method based on low-pass filtering, to
remove the spin-induced modulations. In this section, we check the
robustness of this method by applying it to a set of 50 \sprecing
simulations from the \SXS catalog~\cite{Scheel:2025jct} with $10^{-4}
\lesssim \eSpEC \lesssim 10^{-2}$. A list of these simulations are
provided in table~\ref{tab:sxs_sims}
in~\ref{sec:sxs_simulations_robustness_check_in_small_eccentricity__large_sprec_regime}.

In figure~\ref{fig:small_nr_eccentricity}, we plot the $\egw$ at
$\tStartHat$ vs $\eSpEC$ at $\tRefSpEC$ for this set of simulations,
where $\tRefSpEC$ is the reference time where $\eSpEC$ is measured in
the \SpEC metadata. All of the cases in the figure have $\rFilter >
0.2$, and therefore, we apply the low-pass filter to $\omegaGW$. Note
that the $\egw$ measured using $\omegaGWFiltered$ is consistent with
$\eSpEC$ within a factor of $\sim 2$. The absolute difference is of
the order of $\eSpEC$. On the other hand, the $\egw$ measured using
$\omegaGW$ without applying filtering shows a significant
deviation as $\eSpEC$ goes below $10^{-3}$, and $\egw$ can
differ from $\eSpEC$ by a factor of $\sim 10$. Note that $\eSpEC$ is
computed from the black hole trajectories by fitting the first
derivative of the orbital frequency~\cite{Scheel:2025jct}, whereas
$\egw$ is computed using $\omegaGW$ from the GW waveforms. Also,
$\eSpEC$ is only meant to be an approximate estimate of the
eccentricity, and should be treated only as a
reference~\cite{Scheel:2025jct}. Nevertheless, it serves as a useful
check for the filtering method.

\section{Conclusion}
\label{sec:conclusion}
We present a generalized definition of eccentricity, $\egw$, and mean
anomaly, $\lgw$, for compact binaries on generic bound orbits that
exhibit both eccentricity and \sprec. Spin-precession adds two
complexities in the eccentricity measurement from GW waveforms ---
precession of the orbital plane about the total angular momentum of
the system, and the \masym between positive and negative $m$
modes. The precession of the orbital plane causes leakage of GW power
from the ($\ell=2, m=\pm 2$) modes into ($\ell=2, m \neq 2$). To fix
this, following previous
works~\cite{Estelles:2020osj,Estelles:2020twz,Estelles:2021gvs,Akcay:2020qrj,Gamba:2021ydi,Hamilton:2021pkf,Ramos-Buades:2023ehm,Gamba:2024cvy},
we transform the waveforms to the coprecessing frame, and reestablish
the mode hierarchy where the $(\ell=2, m=\pm 2)$ modes remain the
dominant modes during the entire inspiral. However, even in the
coprecessing frame, the \masym causes nonmonotonicity in the values of
the \dm mode frequency at the extrema. To address this,
following previous
works~\cite{Boyle:2014ioa,Blackman:2017pcm,Blackman:2017dfb,Varma:2019csw},
instead of using only the \dm mode, we use amplitude $\ampGW$ and
phase $\phaseGW$ defined by the symmetric and anti-symmetric
combination of the amplitudes and phases, respectively, of the $(2,
2)$ and $(2, -2)$ modes in the coprecessing frame. We then use
$\ampGW$, $\phaseGW$, and $\omegaGW$ (the first time derivative of
$\phaseGW$) to define the eccentricity $\egw$ and mean anomaly $\lgw$.

To ensure the monotonicity of $\egw(t)$, we also adopted an
interpolation method, \mRatFit, for building the interpolants
$\omegaGWP(t)$ and $\omegaGWA(t)$ for $\omegaGW$ values at the
pericenters and the apocenters, respectively, using rational function
approximation.  We find \mRatFit to be highly robust and better than
the Spline based method, \mSpline, in capturing the trend of
$\egw$ near the merger and avoiding small numerical noise that might
be present in the waveform data.

We find our definition of eccentricity using $\ampGW$, $\phaseGW$ and
$\omegaGW$ to be robust in measuring eccentricity for highly \sprecing
systems. We demonstrated its
robustness by applying it to NR as well as \TEOB waveforms. We also
performed smoothness tests by plotting --- (a) $\egw$ at initial time
vs initial model input eccentricity $\eEOB$, and (b) the evolution of
$\egw$ over time --- for a large set of \sprecing \TEOB waveforms with
the model input eccentricity varying from $\eEOB = 10^{-5}$ to
$\eEOB=0.8$.

For small eccentricity ($\egw \lesssim 10^{-2}$) and high \sprec
($\chi_p \approx 0.8$), spin-induced modulations due to
\masym may become non-negligible compared to eccentric
modulations. We provide robust method based on low-pass
filtering to remove such spin-induced oscillations.  We find that the
spin-induced modulation oscillate at a timescale about 2-3
times the timescale of the eccentric modulations. We utilize this
difference in the timescale of these two effects to build a consistent
strategy for removing the spin-induced modulation by applying a
low-pass filter.

The spin-induced modulations appear only in NR waveforms since the
currently available eccentric waveform
models~\cite{Liu:2023ldr,Gamba:2024cvy,Albanesi:2025txj,Morras:2025nlp}
do not contain \masym. However, in the future, we expect waveform
models to include terms arising from \masym. In that case,
spin-induced oscillations will appear in both $\omegaGW$ and
$\ampGW$. While the method described in
section~\ref{sec:low_eccentricity_and_high_precession} can be used to
remove these oscillations, waveform models will offer a more direct
approach by giving us full control over the \masym
terms. Specifically, we can suppress the spin-induced oscillations by
disabling these terms when generating waveforms for the purpose of
measuring $\egw$. Moreover, such models could be used to remove
spin-induced oscillations from NR waveforms by constructing an
approximate model for these effects. We leave these possibilities for
future work, once eccentric waveform models incorporating \masym become
available.

\texttt{Python} implementation of our methods is publicly available
via package \package.

\ack
The authors thank Keefe Mitman, Rossella Gamba for valuable comments
on the draft, and Aldo Gamboa and Leo C. Stein for useful discussion.
M.A.S.'s research was partially supported by the National Research Foundation of Korea
under grant No.~NRF-2021R1A2C2012473.
V.V.~acknowledges support from National Science Foundation (NSF) Grant
No. PHY-2309301 and UMass Dartmouth's Marine and Undersea Technology (MUST)
Research Program funded by the Office of Naval Research (ONR) under Grant
No. N00014-23-1-2141.
A. Ramos-Buades is supported by the Veni research programme which is
(partly) financed by the Dutch Research Council (NWO) under the grant
VI.Veni.222.396; acknowledges support from the Spanish Agencia Estatal
de Investigación grant PID2022-138626NB-I00 funded by
MICIU/AEI/10.13039/501100011033 and the ERDF/EU; is supported by the
Spanish Ministerio de Ciencia, Innovación y Universidades (Beatriz
Galindo, BG23/00056) and co-financed by Universitat de les Illes Balears.
L. E. Kidder's work was supported by the NSF under Grants No. PHY-2407742,
No. PHY- 2207342, and No. OAC-2209655, and by the Sherman Fairchild Foundation
at Cornell.
This material is based upon work supported by NSF's LIGO Laboratory which is a
major facility fully funded by the NSF.
Part of the numerical calculations reported in this paper, as well
as the development of \package{}, were carried out on the Alice
cluster at the International Centre for Theoretical Sciences, Tata
Institute of Fundamental Research.
\appendix

{
\section{Dependence on waveform extrapolation order}
\label{sec:dependence_on_waveform_extraction_method}
\begin{figure}
\centering
\includegraphics[width=\textwidth]{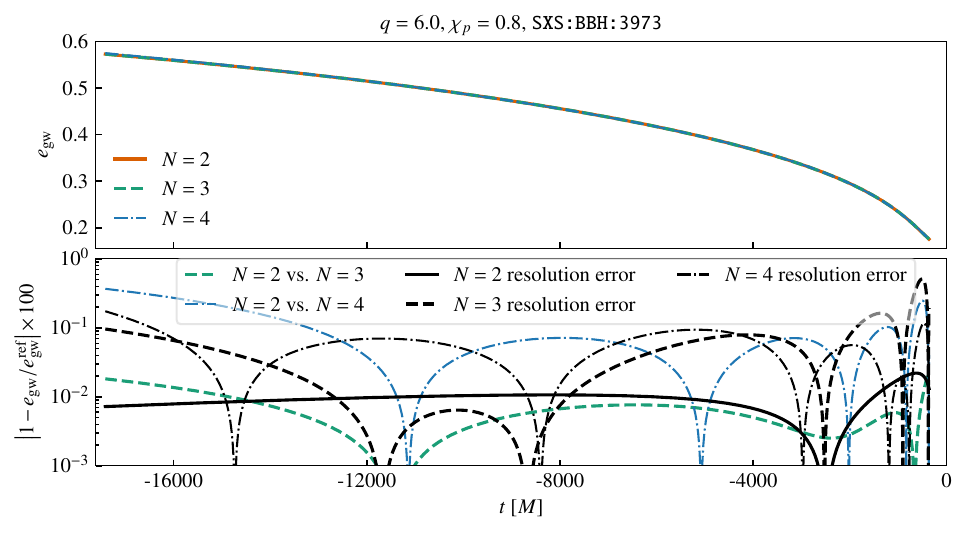}
\label{fig:compare_extrapolation_methods}
\caption{{\bfseries Dependence of $\egw$ on the waveform extrapolation order.}
  {\itshape Top:} $\egw$ computed using waveforms with different extrapolation
order $N$, for the high resolution run of the NR simulation \sxsDemoId.
{\itshape Bottom:} The percentage difference between
$\egw$ and a reference $\egw^{\mathrm{ref}}$.
For the blue and green lines, $\egw$ corresponds to $N \neq 2$ while
$\egw^{\mathrm{ref}}$ corresponds to $N=2$, both for the high
resolution case.
These indicate the
percentage difference in $\egw$ due to different extrapolation orders.
For the black lines, $\egw$ corresponds to medium NR resolution while
$\egw^{\mathrm{ref}}$ corresponds to high NR resolution, both for a fixed $N$. These
indicate the percentage difference in $\egw$ due to NR truncation error.
}
\end{figure}

In this section, we check the dependence of $\egw$ on the {NR} waveform
extrapolation order~\cite{Scheel:2025jct}. Throughout this work, we have used
the $N=2$ extrapolated waveform to compute $\egw$. However, waveforms using a
different extrapolation order may differ slightly from the $N=2$ extrapolated
waveform. To check the dependence of $\egw$ on the waveform extrapolation
order, we consider an NR simulation \sxsDemoId with large eccentricity and large
\sprec, and plot the resulting $\egw$ in
figure~\ref{fig:compare_extrapolation_methods}.  {We consider NR data from the
two highest resolution runs available for this simulation, and denote them as
high and medium resolution.  } The {\itshape top} panel {of
figure~\ref{fig:compare_extrapolation_methods}} shows the $\egw$ evolution
using different extrapolation orders {for the high resolution run}.  
The {\itshape bottom} panel shows the
percentage difference in $\egw$ between (i) different extrapolation orders for
the high resolution run, and (ii) different NR resolutions for a fixed
extrapolation order. 
As both the differences are at the same level ($\lesssim 0.4\%$), we conclude that the difference between the $N=2$ and
$N \neq 2$ extrapolated waveforms is comparable to differences due to NR truncation error.
}

\section{Rational Fit vs Spline}\label{sec:rational_fit}
\begin{figure}
  \centering
  \includegraphics[width=\textwidth]{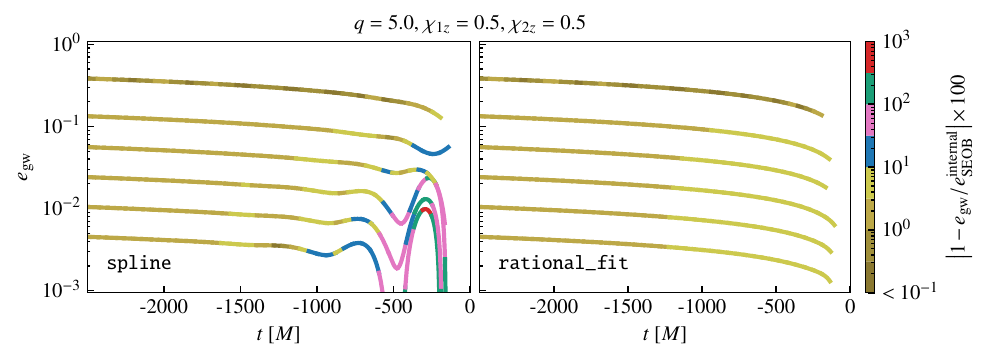}
  \caption{{\bfseries Comparison of $\egw$ measurements using two
interpolation methods for $\omegaGWP(t)$ and $\omegaGWA(t)$.}
{\itshape Left:} $\egw$ computed using the \mSpline interpolants for
$\omegaGWP(t)$ and $\omegaGWA(t)$. {\itshape Right:} presents the same
using the \mRatFit method. In both panels, the colors indicate the
instantaneous percentage difference $\left|1 - \egw/\eSEOB\right|$ between
$\egw$ and $\eSEOB$, where $\eSEOB$ is the eccentricity defined
internally by $\SEOB$. The \mRatFit method demonstrates consistently
reduced oscillations in $\egw$ compared to the \mSpline method. As a
result, the maximum difference between $\eSEOB$ and $\egw$ remains
below 10\% with \mRatFit, whereas it can reach up to $\sim 1000\%$
when using \mSpline, especially for small eccentricities ($\egw
\lesssim 5 \times 10^{-2}$) and near the merger. $\egw(t)$ in this
figure are computed using \mResAmp method.
}
  \label{fig:spline_vs_rational_fit}
\end{figure}
In section~\ref{sec:rational_fits_for_omega_extrema}, we
discussed why \mRatFit is a better choice when building the
$\omegaGWP(t)$ and $\omegaGWA(t)$ interpolants compared to using
\mSpline. In this section, we study the robustness of these two
methods by applying them on a set of \aspin waveforms generated using
\SEOB~\cite{Gamboa:2024imd,Gamboa:2024hli} with eccentricity varying
from $\eEOB \gtrsim 10^{-3}$ to $\eEOB = 0.8$ \footnote{We restrict
the upper limit for $\eEOB$ to 0.8. For $\eEOB \gtrsim 0.8$, we find
$\eSEOB$ to become oscillatory for the initial values used in the
test. See section IV.E in \cite{Gamboa:2024hli} for the domain of
robustness ($\eEOB \lesssim 0.7$) of \SEOB.}. To check the robustness,
we compare the measured $\egw$ to $\eSEOB$ defined within the EOB
dynamics in \SEOB using PN expressions.

In figure~\ref{fig:spline_vs_rational_fit}, we plot $\egw$ using
\mSpline on the {\itshape left} panel and the same using \mRatFit on
the {\itshape right} panel. In both the panels, the colors represent
the instantaneous percentage difference $\left|1 -
\egw/\eSEOB\right|$. Similar to
figure~\ref{fig:interpolation_method_comparison_demo}, the difference
is small for regions far from the merger, and are of the same order
for both \mSpline and \mRatFit. However, close to the merger, \mSpline
shows larger relative difference due to the oscillations introduced by
\mSpline. On the other hand, the $\egw$ is consistently monotonic in
case of \mRatFit. The maximum difference with respect to $\eSEOB$ over
the set of these waveforms is orders of magnitude smaller for \mRatFit
(maximum differences fall below $\sim10\%$) compared to \mSpline
(maximum differences reach up to $\sim 1000\%$), especially for small
eccentricities ($\egw \lesssim 5 \times 10^{-2}$) and near the merger.

\section{Eccentricity measurement in presence of numerical
  noise}\label{sec:eccentricity_measurement_in_presence_of_numerical_noise}
One needs to be careful when measuring eccentricity from waveforms
containing numerical noise, for example, obtained using NR
simulations. Recently it was highlighted in~\cite{Islam:2025oiv}
(e.g., see their figure 3) that when computing $\egw$ with \mSpline on a set of non-spinning eccentric NR waveforms
from~\cite{Hinder:2017sxy}, the $\egw(t)$ curves, for a few cases,
exhibit oscillations. In this section, we argue that the oscillations
showed in~\cite{Islam:2025oiv} are driven mainly by the numerical
noise in the waveforms.

\begin{figure}
  \centering
  \includegraphics[width=\textwidth]{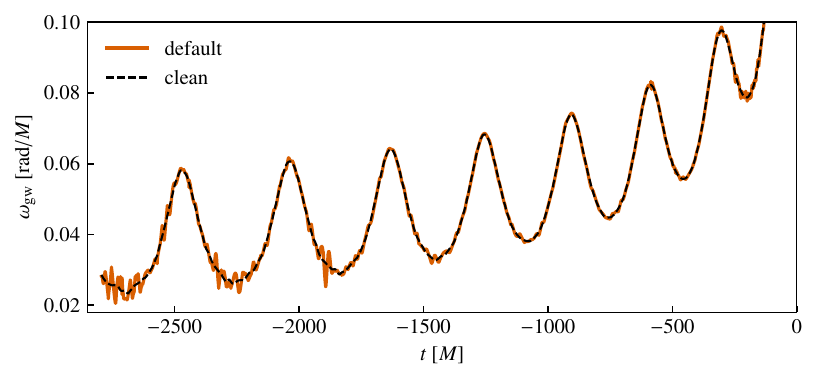}
  \caption{Numerical noise in $\omegaGW$ (solid line) from
    \sxsID{1370}. The dashed line is obtained by applying
    Savitzky-Golay~\cite{Savgol} filter implemented as
    \texttt{savgol\_filter} in \texttt{scipy.signal}.}
  \label{fig:noise_in_omega_gw}
\end{figure}

We consider 19 \SXS waveforms from~\cite{Hinder:2017sxy}. This is the
same set of simulations used in~\cite{Islam:2025oiv}, except that we
have excluded \sxsID{1363}, which has been deprecated in the latest
\SXS catalog~\cite{Scheel:2025jct}, and included \sxsID{1374}, which
was not used in~\cite{Islam:2025oiv}. To illustrate the numerical
noise present in the waveforms, we take a representative simulation
\sxsID{1370} ($q=2$), and plot the corresponding $\omegaGW$ (solid
line, referred to as ``default'') in figure~\ref{fig:noise_in_omega_gw},
where one can see the presence of numerical noise. These numerical
noise in $\omegaGW$ can be removed using a filter such as the
Savitzky-Golay filter~\cite{Savgol}, which is available as
\texttt{savgol\_filter} in \texttt{scipy.signal}. The dashed line
(referred to as ``clean'') in figure~\ref{fig:noise_in_omega_gw} shows
the $\omegaGW$ after removing the noise using \texttt{savgol\_filter}.

\begin{figure}
  \centering
  \includegraphics[width=\textwidth]{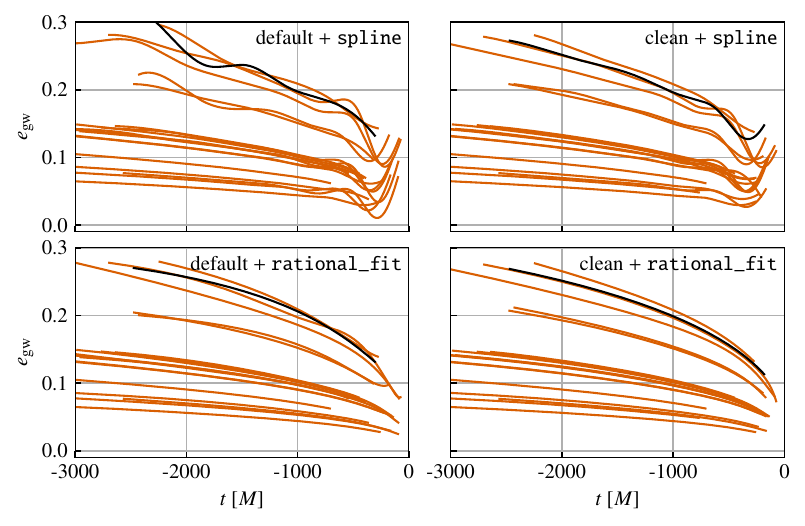}
  \caption{$\egw$ measurement of 19 \SXS waveforms
    from\cite{Hinder:2017sxy}. {\itshape Top-left:} $\egw(t)$ using
    \mSpline without filtering out the numerical noise from the
    waveforms. The numerical noise affect the interpolants built using
    \mSpline. {\itshape Top-right:} $\egw(t)$ using \mSpline after
    filtering out the numerical noise from the waveforms. The 
    oscillations in the early parts, seen in the {\itshape top-left}
    panel, are gone. {\itshape Bottom-right:} $\egw(t)$ using
    \mRatFit without filtering out the numerical noise from the
    waveforms. Oscillations in $\egw(t)$, seen in {\itshape top-left}
    panel, are mostly removed. This is because \mRatFit is less
    sensitive to small numerical noise. {\itshape Bottom-right:}
    $\egw(t)$ using \mRatFit after filtering out the numerical noise
    from the waveforms. $\egw(t)$ does not show any visible
    oscillations. The black curve in each panel represents a case
    (\sxsID{1370}) where the waveform contains numerical noise, as
    shown in figure~\ref{fig:noise_in_omega_gw}.}
  \label{fig:spline_vs_ratfit_sxs_1355_1374}
\end{figure}

In figure~\ref{fig:spline_vs_ratfit_sxs_1355_1374}, we plot $\egw(t)$
for the 19 \SXS waveforms mentioned above. The {\itshape
top-left} panel shows $\egw(t)$ measured using \mSpline, which
exhibits oscillations similar to those reported in
\cite{Islam:2025oiv}. The oscillations in the earlier times are caused
by numerical noise present in the waveforms (as shown in
figure~\ref{fig:noise_in_omega_gw} for a case corresponding to the
$\egw(t)$ highlighted in dark), whereas the oscillations close to
merger are due to limitation of \mSpline method. The fact that the
numerical noise is the main reason behind the oscillations in the
early parts is evident when we apply \texttt{savgol\_filter} to remove
the noise from the 
waveforms. In the {\itshape top-right} panel, we plot $\egw(t)$
with \mSpline after removing numerical noise from the waveforms. Note
that there are no oscillations in the early part of $\egw(t)$. There
are still oscillations near the merger, which we expect when using
\mSpline from our discussion in
section~\ref{sec:rational_fits_for_omega_extrema} and
~\ref{sec:rational_fit}.

Next, we explore how $\egw(t)$ behaves when using \mRatFit instead of
\mSpline. In the {\itshape bottom-left} panel of figure~\ref{fig:spline_vs_ratfit_sxs_1355_1374}, we plot $\egw(t)$ using
\mRatFit without removing the numerical noise from the waveforms. We
note that, unlike $\egw(t)$ in {\itshape top-left} panel where
\mSpline was used, $\egw(t)$ in the {\itshape bottom-left} panel
exhibit significantly lower oscillations. This can be attributed to
the fact that \mRatFit is less sensitive to small numerical noise than
\mSpline. Finally, in the {\itshape bottom-right} panel, we plot
$\egw(t)$ using \mRatFit with the numerical noise removed from the
waveforms, where no visible oscillations are noted.

To summarize, when working with NR waveforms, it is important to
account for numerical noise, especially when working with quantities
like the GW frequency (which is used to define $\egw$).  While using
\mRatFit can help reduce the artificial oscillations in $\egw(t)$
caused by such noise, it may not be sufficient if the noise level is
high. Therefore, for a smooth and monotonic $\egw(t)$, we recommend
both using \mRatFit and applying appropriate noise removal. Since a
one-size-fits-all filtering approach is unlikely to effectively handle
varying levels of numerical noise across different waveforms, we leave
the task of noise removal to the user, but provide diagnostic plots
within \package for quantities such as $\omegaGW$ to guide the
user~\cite{demo_notebook}.

\section{SXS simulations used in figure~\ref{fig:small_nr_eccentricity}}\label{sec:sxs_simulations_robustness_check_in_small_eccentricity__large_sprec_regime}

\begin{table*}[h]
  \caption{\label{tab:sxs_sims}List of simulations from the SXS
catalog~\cite{Scheel:2025jct} used in
figure~\ref{fig:small_nr_eccentricity} for robustness check in the
small eccentricity regime.
}
  \begin{indented}
  \item[]\begin{tabular}{@{}lllll}
           \br          
           \sxsID{0163} & \sxsID{1284} & \sxsID{1344} & \sxsID{3053} & \sxsID{3435} \\
           \sxsID{0316} & \sxsID{1291} & \sxsID{1383} & \sxsID{3153} & \sxsID{3463} \\
           \sxsID{0761} & \sxsID{1298} & \sxsID{1384} & \sxsID{3232} & \sxsID{3516} \\
           \sxsID{0846} & \sxsID{1304} & \sxsID{1867} & \sxsID{3357} & \sxsID{3517} \\
           \sxsID{1004} & \sxsID{1305} & \sxsID{2191} & \sxsID{3358} & \sxsID{3523} \\
           \sxsID{1104} & \sxsID{1314} & \sxsID{2685} & \sxsID{3363} & \sxsID{4037} \\
           \sxsID{1109} & \sxsID{1322} & \sxsID{2811} & \sxsID{3372} & \sxsID{4053} \\
           \sxsID{1239} & \sxsID{1327} & \sxsID{2828} & \sxsID{3380} & \sxsID{4074} \\
           \sxsID{1252} & \sxsID{1329} & \sxsID{2859} & \sxsID{3390} & \sxsID{4153} \\
           \sxsID{1266} & \sxsID{1341} & \sxsID{2970} & \sxsID{3416} & \sxsID{4438} \\
           \br
         \end{tabular}
       \end{indented}
     \end{table*}

\section*{References}
\bibliography{References}

\end{document}